\documentclass[preprint]{aastex}

\usepackage{epsfig}
\usepackage[]{natbib}

\newcommand{\eg}{{\rm e.g.}}
\newcommand{\ie}{{\rm i.e.}}
\newcommand{\Mo}{\ensuremath{M_\odot}}

\newcommand{\Lo}{\ensuremath{L_\odot}}

\newcommand{\Mstar}{\ensuremath{M_{\star}}}

\newcommand{\Lstar}{\ensuremath{L_{\star}}}
\newcommand{\Lstarrat}{\ensuremath{L/L_{\star}}}
\newcommand{\rtLstarrat}{\ensuremath{(L/L_{\star})^{1/2}}}

\newcommand{\rtLeffrat}{\ensuremath{(L_{\rm eff}/L_{\star})^{1/2}}}

\newcommand{\vstar}{\ensuremath{v_{\star}}}
\newcommand{\vstarsq}{\ensuremath{v_{\star}^{2}}}

\newcommand{\phistar}{\ensuremath{\phi_{\star}}}

\newcommand{\amin}{\ensuremath{\arcmin}}
\newcommand{\asec}{\ensuremath{\arcsec}}

\newcommand{\mlb}{\ensuremath{M/L_{B}}}

\newcommand{\kmsMpc}{~\ensuremath{{\rm km\ sec}^{-1}\ {\rm Mpc}^{-1}}}
\newcommand{\kms}{~\ensuremath{{\rm km\ sec}^{-1}}}
\newcommand{\hkpc}{~\ensuremath{{h}^{-1}{\rm kpc}}}

\newcommand{\vnumratsq}{\ensuremath{(v/360 \kms)^{2}}}
\newcommand{\vnumstarratsq}{\ensuremath{(v_{\star}/360 \kms)^{2}}}

\newcommand{\Hnought}{\ensuremath{{\rm H}_0}}
\newcommand{\znought}{\ensuremath{z_0}}
\newcommand{\omeganought}{\ensuremath{\Omega_{0}}}
\newcommand{\lambdanought}{\ensuremath{\lambda_{0}}}
\newcommand{\omegamnought}{\ensuremath{\Omega_{m0}}}
\newcommand{\omegalnought}{\ensuremath{\Omega_{\lambda0}}}

\def \figwidth {\linewidth}

\slugcomment{To appear in Ap.J.}

\shorttitle{Title}
\shortauthors{Wilson \etal}

\begin{document}

\title{GALAXY HALO MASSES FROM GALAXY-GALAXY LENSING\altaffilmark{1}}

\author{Gillian Wilson\altaffilmark{2,3}, Nick Kaiser\altaffilmark{2}, Gerard A. Luppino\altaffilmark{2} and Lennox L. Cowie\altaffilmark{2}}
\email{gillian@het.brown.edu}
\altaffiltext{1}{Based on observations with the Canada-France-Hawaii Telescope which is operated by the National Research Council of
Canada, le Centre National de la Recherche Scientifique de France, and the University of Hawaii.}
\altaffiltext{2}{Institute for Astronomy, University of Hawaii, 2680 Woodlawn Drive, Honolulu, HI 96822}
\altaffiltext{3}{Physics Department, Brown University, 182 Hope Street, Providence, RI 02912}

\begin{abstract}
We present measurements of the extended dark halo profiles 
of bright early type galaxies at redshifts
$0.1 < z < 0.9$  
obtained 
via galaxy-galaxy lensing analysis of
images taken at the CFHT using the UH8K CCD mosaic camera.
Six $0^\circ 5 \times 0^\circ 5$ 
fields were observed for a total of 2 hours each in $I$ and $V$,
resulting in catalogs containing $\sim 20 000$ galaxies per field.
We used $V-I$ color and $I$ magnitude to select bright early type galaxies as the
lens galaxies, yielding a sample of massive lenses with fairly
well determined redshifts
and absolute magnitudes $M \sim M_* \pm 1$.
We paired these with faint galaxies lying at angular distances
$20 \asec < \theta < 60 \asec$, corresponding to
physical radii of $ 26 < r < 77 \hkpc$ ($z = 0.1$) and $105 < r <315 \hkpc$ ($z =
0.9$), and computed the mean tangential shear $\gamma_{T}(\theta)$ of 
the faint galaxies.  
The shear falls off with radius roughly as $\gamma_{T} \propto 1/\theta$ 
as expected for flat rotation curve halos. 
The shear values were weighted in proportion to the square root
of the luminosity of the lens galaxy.  This is optimal if the
halo mass at a given radius varies as $M \propto \sqrt{L}$,
as is the case at smaller radii, and in this context our results
give a value for the average mean rotation velocity of an $L_\star$ galaxy 
halo at $r \sim 50-200\hkpc$ of
$v_\star =  238^{+27}_{-30}\kms$ for a flat lambda 
($\Omega_{{\rm m}0} = 0.3, \Omega_{\lambda 0} = 0.7$) cosmology
($v_\star = 269^{+34}_{-39}\kms$ for Einstein-de Sitter), and with 
little evidence for evolution with redshift.
These halo masses are somewhat ($2-3$ times) 
lower than a simple perfectly flat rotation curve extrapolation 
from smaller-scale dynamical measurements.  They are also
considerably lower than the masses of halos found from the
best studied X-ray halos. They do however agree extremely well with the 
masses of halos of the same abundance in lambda-CDM simulations.
We find a mass-to-light ratio of $\mlb \simeq 121\pm28h(r/100\hkpc)$ (for $\Lstar$ 
galaxies) and these halos
constitute $\Omega \simeq 0.04 \pm 0.01(r/100\hkpc)$ of closure density. 
\end{abstract}


\section{INTRODUCTION}
\label{sec:intro}

The existence of extended dark matter halos with approximately flat rotation curves
around galaxies is now well established.  
At small scales, the halo mass can be measured from
stellar velocity dispersions and
rotation curves, and globular cluster kinematics (\eg\ reviews
by \markcite{fg-77,trim-87}{Faber} \& {Gallagher} 1979; {Trimble} 1987).
Spiral galaxy HI rotation curves \markcite{bos-81}({Bosma} 1981) extend this
and indicate $M \propto r$ out to tens of kpc.  Relative motions of
faint satellites \markcite{bt-81,zar-97}({Bahcall} \& {Tremaine} 1981; {Zaritsky} {et~al.} 1997)
or
pairs of galaxies \markcite{tur-76,jmb-98}({Turner} 1976; {Jing}, {Mo}, \& {Boerner} 1998)
analyzed statistically
extend this to larger scales, and at still larger scales the
cosmic virial theorem analysis \markcite{dp-83}({Davis} \& {Peebles} 1983)
shows that relative motions remain flat or
slowly rising to scales of a few Mpc, suggesting that the average mass around 
a galaxy continues to rise roughly in proportion to radius.
Galaxy clustering measurements show that the excess {\sl light}
around a galaxy is $L_{\rm excess}(<r) \simeq 4 \pi \xi(r) \mathcal{L} r^{3}$
where $\cal{L}$ is the mean luminosity density. 
This also grows roughly in proportion to radius.  The excess light is
equal to $L_\star$ at a radius of $r \sim 400 h^{-1}$kpc. On 
scales larger than this one is dealing 
not with individual halos but with 
the collective mass of collections of neighboring galaxies.  
Here we shall restrict attention
to smaller scales where it is reasonable to interpret the results
as probing relatively stable and virialized halos of individual
galaxies.

The halos of early type galaxies can also be probed 
via X-ray 
emitting hot gas.  This is valuable as it removes some of the
uncertainty regarding orbital anisotropy in above analyses.  Unfortunately
the halos are very faint, and only a handful of galaxies have the resolved flux and
temperature data required \markcite{kf-95, trinch-97}({Kim} \& {Fabbiano} 1995; {Trinchieri}, {Fabbiano}, \&  {Kim} 1997).  In the best studied case (NGC 4636) 
\markcite{mush-94, trinch-94}({Mushotzky} {et~al.} 1994; {Trinchieri} {et~al.} 1994) the halo is very
massive indeed $M(< 100 {\rm kpc}) \simeq 5.1\times 10^{12} M_\odot$. 
The line of sight stellar velocity dispersion for this galaxy is 
$\sigma \simeq 191\kms$ ---
comparable to the mean value $\sim 210\kms$ for $L_\star$ galaxies
\markcite{ft-91}(Fukugita \& Turner 1991) --- corresponding to a rotation
velocity $\sim 330\kms$ for the luminous region, whereas the
X-ray mass at $100$ kpc gives a rotation velocity of $470\kms$.

The mass of galaxy halos at radii $\sim 100-300 $ kpc is of considerable
importance both in the accounting of the matter content of the universe and
in testing cosmological theories (which are typically finely tuned to match
the properties of massive galaxy clusters).
However, the dynamical measurements suffer from systematic modeling
uncertainties, and it is difficult to know whether halos like that
of NGC 4636 are typical of ordinary bright ellipticals.

Gravitational lensing offers an alternative probe of the dark matter around galaxies.
The manifestation of lensing which we shall exploit here is the weak `galaxy-galaxy lensing'
effect; the distortion of shapes of (typically faint)
background galaxies seen near (typically brighter) foreground galaxies.
Clusters of galaxies have traditionally been the primary target of weak
lensing studies (see \markcite{mel-99}{Mellier} (1999) for a review). Individual  galaxy masses are far more difficult to measure
due to their being less massive and hence yielding a smaller lensing
signal relative to the noise. However, by stacking pairs of galaxies it is possible to 
beat down the noise and measure the total average halo profile.

In galaxy-galaxy lensing one measures the
the mean tangential shear of faint `source' galaxies averaged over source-lens pairs
binned by angular separation:
\begin{equation}
\label{eq:gammaTdef}
\gamma_{T}(\theta) = - {
\sum\limits_{\rm pairs} W_l W_s M_{\alpha ij} \theta_i \theta_j \hat{\gamma}_\alpha / \theta^2
\over \sum\limits_{\rm pairs}  W_l W_s}
\end{equation}
where $\hat{\gamma}_\alpha$, for $\alpha = 1,2 $, is the shear estimate for the source galaxy,
$\theta$ is the projected 
angular separation of the lens and source,
$W_l$, $W_s$ are weights for the lens and source,
and the two constant
matrices $M_1$, $M_2$ are 
\begin{equation}
\label{eq:matrices}
M_{1lm} \equiv \left[ 
\begin{array}{cc}
1 & 0\\
0 & -1\\
\end{array}\right],
\;
M_{2lm} \equiv \left[ 
\begin{array}{cc}
0 & 1\\
1 & 0\\
\end{array}\right].
\end{equation} 

The expectation value of the mean tangential shear is related to the mean excess dimensionless
mass surface density $\kappa(\theta)$ by
\begin{equation}
\label{eq:gammatfromkappabar}
\langle \gamma_{T}(\theta) \rangle = - 
{\partial \overline{\kappa} \over \partial \ln \theta^2}
\end{equation}
with
\begin{equation}
\label{eq:kappabarfromkappa}
\overline{\kappa}(\theta) \equiv {1 \over \pi \theta^2} \int_{|\theta'| < \theta}   d^2 \theta' \; \kappa({\theta'}) 
\end{equation}
\markcite{ksfw-95}(Kaiser {et~al.} 1995).
The dimensionless excess surface density is in turn related to the
galaxy-mass cross correlation function $\xi_{g\rho}$ by
\begin{equation}
\label{eq:kappafromwgrho}
\kappa(\theta) = {4 \pi G \over c^2} 
{\int dz_l \; n_l(z_l) D_l \langle \beta (z_{l}) \rangle
\int dM_L \; \Phi(M_l; z_l) \langle W_l(M_l, z_l) \rangle  
\int dy \; \xi_{g\rho}(\sqrt{y^2 + D_l^2 \theta^2}; M_l, z_l) 
\over \int dz_l \; n_l(z_l) \int dM_l\; \Phi(M_l; z_l) \langle W_l(M_l, z_l) \rangle}
\end{equation}
where $n_l(z)$ is the redshift distribution for the lens galaxies;
$\Phi(M; z)$ is the absolute magnitude distribution at redshift $z$;
$\xi_{g\rho}(r; M, z)$ is the galaxy-mass cross correlation;
defined as the mean physical density at a physical distance $r$ from
a lens galaxy, parameterized by the absolute magnitude and redshift.
The quantity $\langle W_l(M, z) \rangle$ is the mean weight for
galaxies of a given absolute magnitude and redshift.  
The angular diameter distance is
$D_l \equiv a_0 \omega_l / (1 + z_l)$ 
where $\omega$ is comoving distance measured in units of the current curvature scale
$a_0 = c / (H_0 \sqrt{1 - \omegamnought - \omegalnought})$.
The dimensionless quantity $\langle \beta (z_{l}) \rangle$ is
defined as 
\begin{equation}
\label{eq:betaeff}
\langle \beta (z_{l}) \rangle \equiv \frac{\int^{\infty}_{0} dz_s \; n_s(z_s) 
\langle W_s(z_s) \rangle \beta(z_l, z_s)}
{\int^{\infty}_{0} dz_s \; n_s(z_s) \langle W_s(z_s) \rangle}
\end{equation}
where $n_s(z)$ is the redshift distribution of the source galaxies,
$\langle W_s(z_s) \rangle$ is the mean weight for source galaxies at
redshift $z_s$,
and where, finally,
\begin{equation}
\label{eq:betalambda}
\beta(z_l, z_s) \equiv {\rm max}(0,\sinh{(\omega_{s}-\omega_{l})}/\sinh{(\omega_{s})}) 
\end{equation}
Physically, $\beta(z_l, z_s)$ is the ratio of the distortion induced by
a lens at redshift $z_l$ in
an object at
finite distance $\omega(z_s)$ relative to that for a fictitious source at infinite
distance.

For the special case of a spatially flat cosmology, $\omega \rightarrow 0$ and
$a_0 \rightarrow \infty$, but such that their product remains finite.  In that case
$\sinh \omega \rightarrow \omega$, and
$\langle \beta \rangle \equiv  \langle {\rm max}(0,1 - \omega_{l}/\omega_{s})\rangle$.
For the limiting case of $\Omega_m = 1$, $\Omega_\lambda = 0$,
$\omega(z) = 2(1 - 1/ \sqrt{1 + z})$ and, 
in the other extreme, for  $\Omega_m \rightarrow 0$, $\Omega_\lambda \rightarrow 1$,
$\omega(z) = z$.

Equations (\ref{eq:betalambda}, \ref{eq:kappafromwgrho}, \ref{eq:kappabarfromkappa},
\ref{eq:gammatfromkappabar})
provide a direct relationship between observable $\langle\gamma_{T}\rangle$ 
and the cosmologically interesting quantity $\xi_{g\rho}$.  They allow one to
compute the expected tangential shear given 
a cosmological model, a theoretical $\xi_{g\rho}$,
measured redshift distributions $n_s(z)$, $n_l(z)$ for source and lens galaxies,
and user supplied weights.  The latter should ideally be determined from the image quality
for the sources and from the brightness of the lens galaxies in such a way as
to maximize the signal-to-noise, but the results above are valid for
arbitrary weights.

One can also calculate the variance in the mean tangential shear, and combining
this with the formalism above yields the expected signal-to-noise
ratio.  This exercise shows that the S/N is rather poor
if galaxies are divided into lens and source samples solely on magnitude. 
This
is because the range of redshift at a given apparent magnitude is large, so there is a 
large variation in bright galaxy absolute 
luminosity and therefore in the mass. There is 
also a large range in $\beta$ values. Photometric redshifts are useful in this
regard to tighten up the distribution of foreground lenses, and allow one
to boost the signal to noise by giving weight preferentially to the more massive
galaxies.  

The above equations are quite general.  For the special case of a power-law
galaxy-mass correlation function $\xi_{g\rho}(r) \propto r^{-\gamma}$ then
$\langle \gamma_T(\theta) \rangle \propto \theta^{1 - \gamma}$ with a
constant of proportionality which is computable from the lens, source
redshift distributions etc.  Specializing further, 
for a flat rotation curve object
the shear is given by
\begin{equation}
\label{eq:gammat}
\gamma_{T}(\theta) = \pi (v/c)^2 \langle \beta(z_l) \rangle / \theta 
= 0.93 (v / 360 \kms)^2 (1'' / \theta) \langle \beta(z_l) \rangle
\end{equation}
This equation provides a convenient rule of thumb to convert 
between measured shear values and an equivalent rotation velocity.
Similarly, if $\xi_{g\rho} \propto r^{-2}$ we can characterize the
mean halo profile in terms of an equivalent mean rotation velocity, which
is convenient when one comes to compare with dynamical measurements
on smaller scales.
(The fiducial rotation velocity of $360\kms$ is that obtained for an $\Lstar$ early type galaxy
from Faber-Jackson measurements \markcite{ft-91}(Fukugita \& Turner 1991)). We return to discuss the 
value of rotation velocity as measured by a variety of techniques in \S\ref{sec:disc}. 

Early photographic measurements \markcite{tyson-84}(Tyson {et~al.} 1984) gave an essentially
null detection of tangential shear  which seemed to rule out
extended massive halos, though the upper limit was subsequently revised
upwards \markcite{km-87, kaiswk-91}({Kovner} \& {Milgrom} 1987; Kaiser 1991). The first detection
of galaxy-galaxy lensing was by \markcite{brain-96}Brainerd, Blandford, \& Smail (1996). Since then, a
number of groups have presented estimates of galaxy-galaxy lensing, either 
from Hubble Deep Field observations \markcite{datys-96, griff-96, hud-98}(Dell'Antonio \& Tyson 1996; Griffiths {et~al.} 1996; Hudson {et~al.} 1998), from observations
of the rich cluster AC 114 \markcite{nkse-98}({Natarajan} {et~al.} 1998), 
or from the Sloan Digital Sky Survey \markcite{fisch-00}({Fischer} {et~al.} 2000). These
results demonstrate the practicality of the approach, but there
are some uncertainties concerning their calibration.

In this paper we investigate galaxy-galaxy lensing using data collected at the CFHT
with the UH8K camera.  Our analysis differs somewhat from other works in that
we focus on bright early type galaxy halos, as these are the only class of
objects whose redshifts can be reliably determined from 2-passband photometry.
However, while early type galaxies contribute only $30-50\%$ to the total
luminosity density, dynamical studies of the local universe show that an $L_{\star B}$ elliptical has about four times the mass
(at a given radius) as an $L_{\star B}$ spiral \markcite{ft-91}(Fukugita \& Turner 1991), and 
these objects are expected to dominate the lensing signal at all redshifts.

The outline of the paper is as follows.  In \S\ref{sec:data} we describe the
data and the selection of lens and background galaxies. 
In \S\ref{sec:results} we present tangential
shear measurements for lens galaxies over a wide range of redshifts. 
To facilitate the comparison with other studies and with
predictions from simulations we characterize the halo profiles
in terms of the equivalent rotation velocity for an $L_\star$ galaxy. 
In \S\ref{sec:disc} we discuss our results.  We 
calculate the mass-to-light ratio of an $\Lstar$ early type galaxy and the
contribution of early types to the closure density. 
We also compare our values to other lensing studies, X-ray
measurements,  
and
to the masses of simulated halos of the same abundance.
In \S\ref{sec:conc}  we briefly summarize our conclusions.
We assume a 
flat lambda ($\Omega_{{\rm m}0} = 0.3, \Omega_{\lambda 0} = 0.7$) 
cosmology with $\Hnought = 100$ h$ \kmsMpc $ throughout unless
explicitly stated otherwise. 

\section{THE DATA AND GALAXY SAMPLES}
\label{sec:data}

\subsection{Data Acquisition and Reduction}

The data were taken at the 3.6m CFHT telescope using the $8192 \times 8192$
pixel UH8K camera at prime focus. The field of view of this camera is 
$\sim 30 \amin$ with pixelsize $0.207$\asec. The data (six pointings) used in the analysis were acquired as part
of an ongoing 
project whose principle aim is to investigate the cosmic shear pattern caused
by gravitational lensing from the large-scale structure  of the Universe.
Table~\ref{tab:fields} gives an overview of the data, describing the field 
name, center and seeing for each pointing. This is the second in a series of papers describing results from that project.
\markcite{kwl-01}Kaiser, Wilson, \&  Luppino (2001a, Paper I) presented estimates of cosmic shear variance 
on $2' - 30'$ scales. 
Here we focus on properties of massive galaxy halos at radii of $50 -200\hkpc$.
Forthcoming papers will address galaxy
clustering, and correlations between mass and light on cluster and group 
scales \markcite{wkl-01}(Wilson, Kaiser, \&  Luppino 2001b, Paper III).  A full description
of our catalogs will be presented in a later paper 
\markcite{wkcats-01}(Wilson \& Kaiser 2001). Further details of the data reduction pipeline may be found in \markcite{kwld-01}Kaiser {et~al.} (2001b),
and an application to the ms0302 supercluster in \markcite{kwlkgmd-01}Kaiser {et~al.} (2001c).
In brief, the data was dark subtracted, flat-fielded, registered and median averaged.
Weighted second moment shapes and magnitudes of objects were measured using varying aperture photometry 
and
optimally weighted shear estimates for each galaxy,  $\gamma_{\alpha}$, were determined using the 
method described in \markcite{kais-00}{Kaiser} (2000).

\begin{deluxetable}{ccccrrcc}
\tablewidth{0pt}
\tablecaption{Field Centers and  Seeing 
\label{tab:fields}
}
\tablehead{
\colhead{Field} &
\colhead{Pointing} &
\colhead{RA (J2000)} &
\colhead{DEC (J2000)} &	
\colhead{l } &	
\colhead{b } &	
\colhead{FWHM(I)} &	
\colhead{FWHM(V)} 	
}
\startdata
Lockman	& 1	& 10:52:43.0	& 57:28:48.0 	& $149.28$	& $53.15$	& $0''.83$	& $0''.85$	\\
	& 2	& 10:56:43.0	& 58:28:48.0 	& $147.47$	& $52.83$ 	& $0''.84$	& $0''.86$ 	\\
Groth	& 1	& 14:16:46.0	& 52:30:12.0	& $96.60$	& $60.04$	& $0''.80$	& $0''.93$	\\
	& 3	& 14:09:00.0	& 51:30:00.0 	& $97.19$	& $61.57$	& $0''.70$	& $0''.85$	\\
1650	& 1	& 16:51:49.0	& 34:55:02.0 	& $57.37$	& $38.67$	& $0''.82$	& $0''.85$	\\
	& 3	& 16:56:00.0	& 35:45:00.0 	& $58.58$	& $37.95$	& $0''.85$	& $0''.72$	\\
\enddata
\end{deluxetable}

\subsection{Lens Galaxy Sample}
\label{ssec:fg}

Our analysis differs from other groups in that we use $V -I$ color
to select a sample of bright early type lens galaxies with reasonably
well determined redshifts.  This allows us to focus on a single type
of galaxy --- though it obviously precludes drawing any useful
conclusions about later type galaxies --- and may allow
useful constraints on the evolution of halos over time and
on the profile.  

To a first approximation, galaxies can be divided into 
spectral classes within which the galaxies
have very similar, and largely
luminosity independent, spectral energy distributions (SEDs). For each
type, $t$, there is a color-redshift relation $c = c_t(z)$. With
measurements of two colors (\ie\ a minimum of 3 passbands)
it should then generally be possible to determine both the spectral type and
redshift.  Here we have only fluxes in 2 passbands, but this
is still sufficient to select a subset of galaxies --- bright early
types --- and assign them approximate redshifts.  This is because
early type galaxies are the reddest galaxies at a given redshift.  Thus, if
we select galaxies of some color $c$ we will see a superposition
of early types at redshift $z_E$ such that $c = c_E(z_E)$
and later types at their appropriate, but considerably higher, redshift.  
An $L \sim L_\star$ early
type galaxy 
will appear much brighter than an $L \sim L_\star$ spiral galaxy, 
as we will see shortly, by about 3 magnitudes,
so with
a judicious cut in red flux it should be possible to isolate a 
bright --- and therefore presumably massive --- early type
galaxy sample.
To substantiate these comments we first compute the expected contribution to the
counts as a function of $I$-magnitude for slices in color from galaxies of 
various types using the
local 2dF LF determination assuming no evolution.  We compare these with our observed counts.
We then test the technique
with real high redshift galaxies of Cowie \markcite{cghskw-94,cshc-96, csb-99, wcbb-01}(Cowie {et~al.} 1994; {Cowie} {et~al.} 1996; Cowie, Songaila, \& Barger 1999; Wilson {et~al.} 2001a), and demonstrate the
photometric redshift precision. 

Given the SED $f_\nu$ for galaxies of type $t$, one can compute the color as
a function of redshift $c_t(z)$.  A narrow band of color of width
$dc$ around some color $c$ then corresponds, for that type, to
a range of redshift $dz = (dc_t/dz)^{-1} dc$ around $z = z_t(c)$, this
being the inverse function defined such that $c_t(z_t(c)) = c$.
If the color-redshift curve $c_t(z)$ is non-monotonic then the
inverse function $z_t(c)$ will be multi-valued.
We define the type-specific luminosity function 
$\phi_t(L)$ such that the number of
galaxies of type $t$ in comoving volume $d^3 r$ and in an interval of width
$dL$ around $L$ is
\begin{equation}
\label{eq:dnfromphi}
dn_t = \phi_t(L) dL d^3 r.
\end{equation}
Equivalently, the distribution over absolute magnitude,
most often quoted in terms of $B$-magnitudes, is
\begin{equation}
\label{eq:dnfromPhi}
dn_t = \Phi_{t} (M_B) dM_B d^3 r
\end{equation}
with
\begin{equation}
\Phi_{t}(M_B) \equiv 0.4 \ln(10) L \phi_t(L).
\end{equation}

The apparent magnitude in the $I$-band is 
\begin{equation}
m_I = M_B + 5 \log(D_l(z)/10{\rm pc}) + K_{BIt}(z)
\end{equation}
where $D_l(z)$ is the luminosity distance and where
$K_{BIt}(z) = K_{It}(z) - (M_B - M_I)_{t0}$ is the combination of the
conventional $K$-correction (for galaxy type $t$ in the $I$-band)
and the rest frame color for that type.
At fixed color (and therefore
fixed redshift) $dm_I = dM_B$, while the
comoving volume element is
\begin{equation}
d^3 r = D^2 d\Omega dz {dr \over dz}
\end{equation}
where $d\Omega$ is the solid angle,
$dr$ is a comoving radial distance element, and
$D = D_l / (1 + z)$ is the transverse comoving distance.
The contribution to the counts from galaxies of type $t$ 
and in an infinitesimal range of color $dc$ is then,
from (\ref{eq:dnfromPhi}),
\begin{equation}
\label{eq:colorcounts0}
{dn_t \over d\Omega dm_I} = 
dc D^2 {dr\over dz} \left(dc_t \over dz \right)^{-1}
\Phi_t(m_I - 5 \log(D_l(z_t(c))/10{\rm pc}) - K_{BIt}(z_t(c))).
\end{equation}
In this model --- a universal and
non-evolving SED for each type --- the counts
at a given color are simply a superposition of scaled and shifted
replicas of the various $\Phi_t(M_B)$.  The counts for a
finite range of color $c_1 < c < c_2$ are obtained by
integrating (\ref{eq:colorcounts0}) to give
\begin{equation}
\label{eq:colorcounts1}
{dn_t \over d\Omega dm_I} = 
\int\limits_{c_1 < c_t(z) < c_2} 
 dz\; D^2 (dr / dz) 
\Phi_t(m_I - 5 \log(D_l(z)/10{\rm pc}) - K_{BIt}(z)) 
\end{equation}
which can readily be computed as a discrete sum given
tabulated colors, $K$-corrections as a function of redshift. 

\begin{figure}
\centering\epsfig{file=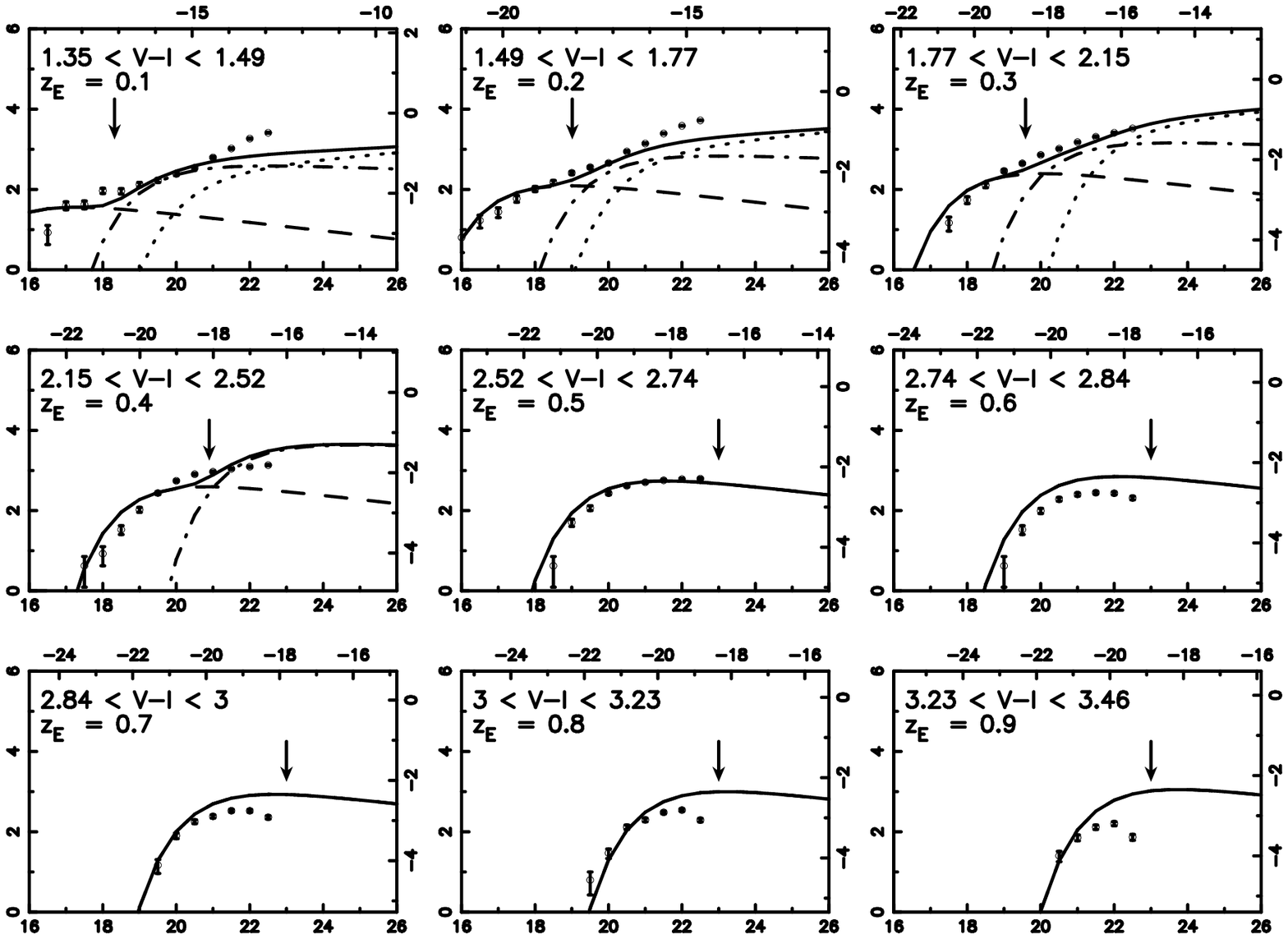,width=\figwidth}
\caption[ctsfancy_FL.ps]{
The symbols with error bars are the log
counts (number of galaxies degree$^{-2}$ mag$^{-1}$) versus 
apparent $I$ magnitude for galaxies in the 
color-ranges indicated. Also shown are predictions for the contribution to the counts
for E/SO (dash), Sbc (dot-dash), Scd (dotted) 
and cumulative (solid) galaxy 
types according to equation~(\ref{eq:colorcounts1}) and with
Schechter function model parameters determined from the 2dF
redshift survey \markcite{folkes-99}({Folkes} {et~al.} 1999)
The upper and right axes apply to early types only and 
show absolute $B$ magnitude and
luminosity function (number of galaxies (h$^{-1}$Mpc)$^{-3}$ mag$^{-1}$). 
The lines in these figures show that for colors corresponding to
moderate redshift ellipticals ($0.1 \lesssim z \lesssim 0.4$ say)
an $L_\star$ elliptical appears $2-3$ magnitudes brighter than an
$L_\star$ spiral.  Thus, by means of a suitable cut in magnitude --- the
value we have adopted is indicated by the arrow --- one can
isolate an essentially pure early type sample.  The $c(z)$ for
spirals peaks at $z \simeq 1$ with $c \simeq 2.4$, and declines for
higher $z$.  This color corresponds to $z_E \simeq 0.4$, and so 
for redder color there are no spirals.  The good agreement
between the predicted and observed counts in the elliptical
dominated regime argues for little evolution of these galaxies.
(There may be some disagreement
with predictions at the highest redshifts due to
slight evolution in $\Lstar$ and/or some additional star formation blueing 
relative to non-evolving predictions). }
\label{fig:lf}\end{figure}

The counts for a set of narrow slices in color (chosen to
correspond to a set of uniform width slices in redshift for
early type galaxies) are shown in Figure~\ref{fig:lf} (lower axis is apparent $I$-magnitude
and left axis is log counts ($\ie$ number of galaxies per square degree per magnitude)) .
Also shown are predictions for the contribution to the counts
according to~(\ref{eq:colorcounts1})
for various galaxy types according the Schechter function model 
for local type-specific 
luminosity function
\begin{equation}
\phi_t(L) dL = \phi_{\star t} 
\left(L \over L_{\star t}\right)^{\alpha_t} e^{-L/L_{\star t}}  
{dL \over L_{\star t}}
\end{equation}
or equivalently
\begin{equation}
\Phi_{t}(M_B) \equiv 0.4 \ln(10)  \phi_{\star t} 
10^{0.4 {(1 + \alpha_t)} (M_{B \star t} - M_B)} 
\exp \left(-10^{0.4 (M_{B \star t} - M_B)} \right).
\label{eq:PhiM}
\end{equation}
with parameters $\phi_{\star t}$, $\alpha_t$ and $M_{B \star t}$ (Table~\ref{tab:2dF})
determined from the 2dF redshift survey by \markcite{folkes-99}{Folkes} {et~al.} (1999) 
and with colors, $K$-corrections
etc, computed using transmission functions for the UH8K system and SED's from \markcite{cole-80}Coleman, Wu, \& Weedman (1980) (2dF types Sab and types
Sbc are combined into one group as $K$-corrections for type Sab are unavailable from CWW).

These plots show that there is very good agreement between
the predicted and observed counts  
The plots also show that the brightest
galaxies at any given color are indeed overwhelmingly dominated
by early type galaxies, so with a cut in apparent $I$-magnitude
indicated by the arrow
it should be possible to isolate a pure early type subsample.
The number of lens 
galaxies selected in each redshift interval (summed over all six pointing) 
using this magnitude cut
may be found in Table~\ref{tab:data}. 
(In Figure~\ref{fig:lf} the upper and right axes refer only to the early type subsample. They
show absolute $B$ magnitude and luminosity function $\ie$ 
number of galaxies (h$^{-1}$Mpc)$^{-3}$ mag$^{-1}$).

\begin{deluxetable}{lccl}
\tablewidth{0pt}
\tablecaption{2dF Schechter Function Fits by Spectral Type.
\label{tab:2dF}
}
\tablewidth{0pt}
\tablehead{
\colhead{Type} &
\colhead{$\Mstar^{B}$} &
\colhead{$\alpha$} &
\colhead{$\phistar (\times10^{-3}(h^{-1}$Mpc)$^{-3})$ }
}
\startdata
${\rm E/S0}$ & $-19.61$  & $-0.740$ & $9.0$
\\
${\rm Sac}$  & $-19.53$  & $-0.925$ & $9.2$
\\
${\rm Scd}$  & $-19.00$   & $-1.210$  & $6.5$
\\
\enddata
\end{deluxetable}

\begin{deluxetable}{crr}
\tablewidth{0pt}
\tablecaption{Data.
\label{tab:data}
}
\tablewidth{0pt}
\tablehead{
\colhead{Lens Redshift} &
\colhead{Number Lens} &
\colhead{Number Pairs}
}
\startdata
$0.1\pm0.05$  & $92$   & $30896$
\\
$0.2\pm0.05$  & $222$  & $78928$
\\
$0.3\pm0.05$  & $366$  & $128533$
\\
$0.4\pm0.05$  & $960$  & $341541$
\\
$0.5\pm0.05$  & $1611$ & $580021$
\\
$0.6\pm0.05$  & $663$  & $237522$
\\
$0.7\pm0.05$  & $699$  & $255391$
\\
$0.8\pm0.05$  & $594$  & $216621$
\\
$0.9\pm0.05$  & $233$  & $84628$
\\
\\
$0.2\pm0.15$  & $680$  & $238357$
\\
$0.5\pm0.15$  & $3234$ & $1159084$
\\
$0.8\pm0.15$  & $1526$ & $556640$
\\
\\
$0.5\pm0.25$  & $4299$ & $1543008$
\\
\enddata
\end{deluxetable}

\begin{figure}
\centering\epsfig{file=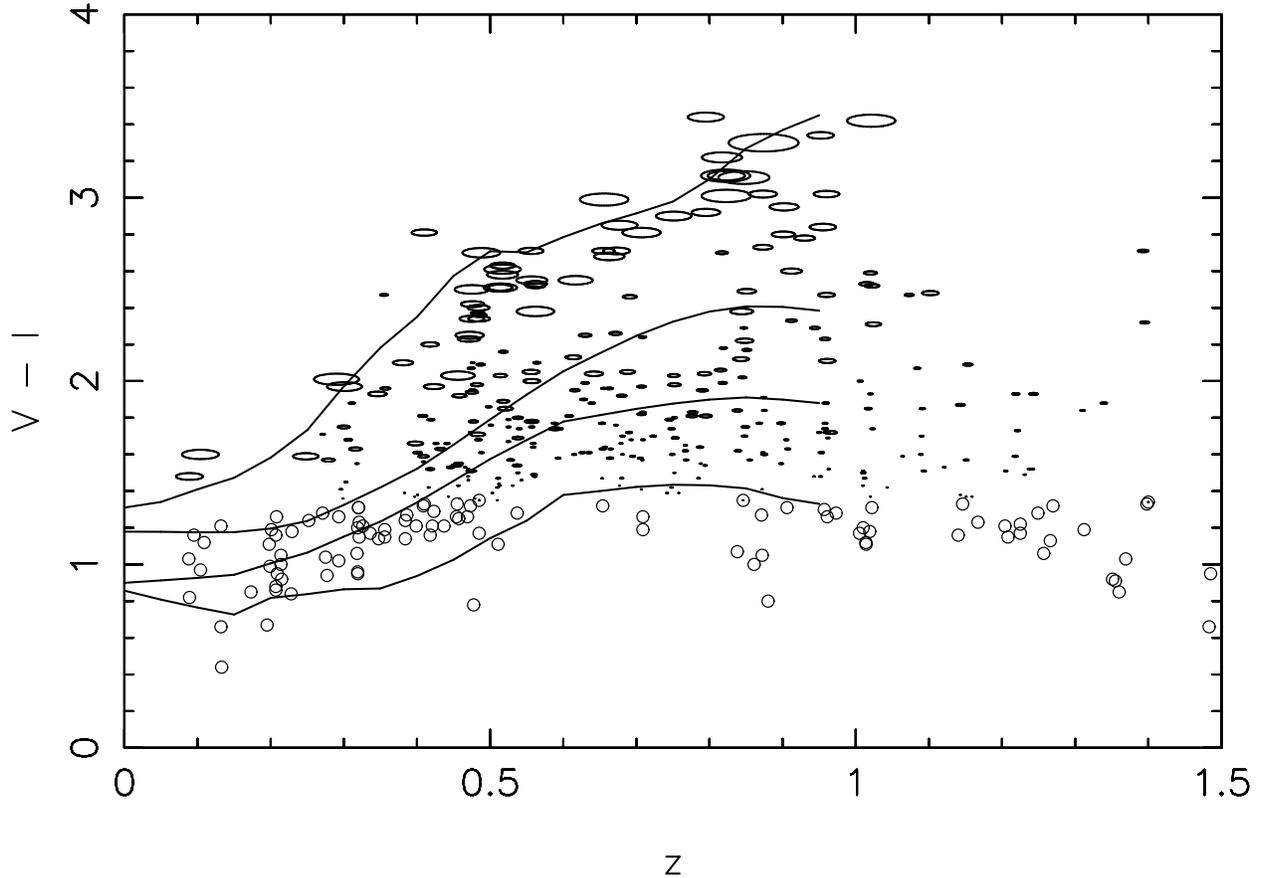,width=\figwidth}
\caption[esel_cz.ps] {
The lines show color {\it vs\/} redshift computed from the \markcite{cole-80}Coleman {et~al.} (1980) SEDs.  
From reddest (top) 
to bluest the lines represent E/SO, Sbc, Scd and Im galaxies.
Symbols are measured colors and spectroscopic redshifts
from Cowie.  The sizes of the ellipses are proportional to the rest-frame absolute B-luminosity of the
galaxy computed using redshift $z_E(c)$ derived from the color assuming an early type SED,
and with $K$-correction $K_E(z_E(c))$ for an early type at that color-redshift.  
The circles indicate galaxies which are bluer than a zero redshift elliptical.
The key point here is that the symbol size is determined entirely from the
broad-band $I,V$ colors, without any reference to the spectroscopic redshift.
These show quite vividly that by selecting galaxies on this property
one obtains a sample of galaxies which are a) with great probability
early type galaxies, and b) have a very tight color redshift relation
$c(z) \simeq c_E(z)$.  The figure also reveals some minor, but interesting,
discrepancies.  There is a well defined sequence of relatively blue galaxies
at $z \sim 0.4 - 0.7$ which seem to track the spiral sequences, but lie
$\simeq  0.2 $ magnitudes below the non-evolving spiral $c(z)$ prediction
This is probably due to evolution, but may also reflect in part some slight differences
between the transmission functions for the standard system filters and those actually used 
at Keck and CFHT.  Also, the redder galaxies at $z \sim 0.5 - 1.0$ again
seem to be slightly bluer than the non-evolving predictions, and this
results in a slight offset in the redshifts determined from the color.
}
\label{fig:esel_cz}
\end{figure}

We can test the accuracy of these photometric redshifts
 using deep redshift surveys.
Figure \ref{fig:esel_cz} shows the $V-I$ colors of Cowie's
sample versus spectroscopic redshift.  Superposed are the
color-redshift curves for the CWW SEDs.
For red galaxies with $c > c_E(z=0)$ the area of the symbol is proportional
to the rest-frame $B$-band luminosity computed from the photometric
redshift $z = z_E(c)$ and $K$-correction $K = K_E(z_E(c))$. 
This shows that the brightest galaxies at any given color
do indeed lie along the upper envelope in color-redshift space
delineated by the early type locus.

\begin{figure}
\centering\epsfig{file=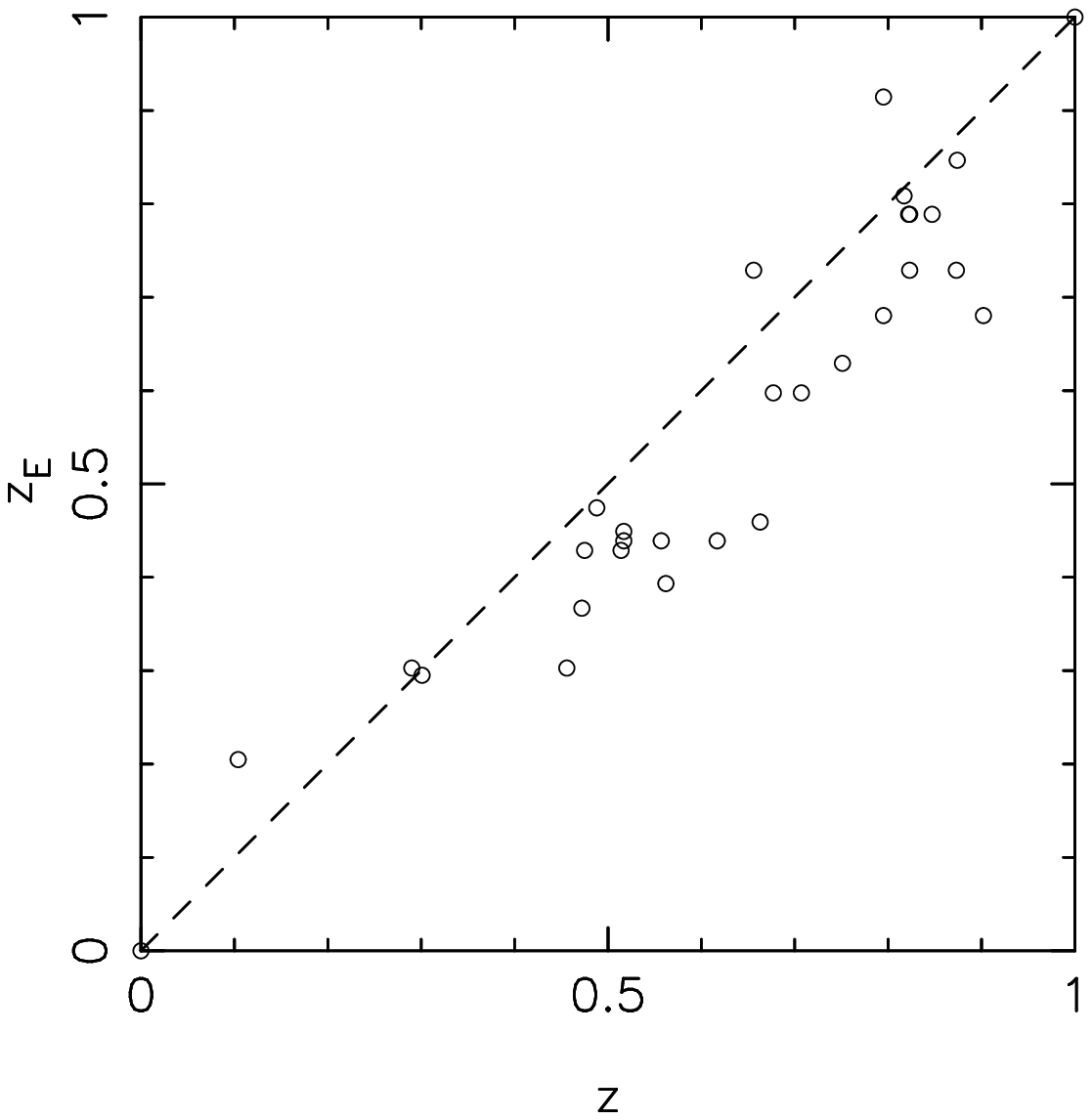,width=\figwidth}
\caption[esel_zz.ps] {
Photometric redshifts $z_E$ for Cowie galaxies derived from $V-I$ color assuming unevolving early type SED.
Only objects with $M_E < -18.6$ (one magnitude fainter than $M_\star$) are shown.
There is very good agreement between these 1-color photometric redshifts
and the spectroscopic results.  The color derived redshifts --- which assume there
has been no evolution of the spectral energy distribution --- appear to be
systematically slightly low.  This shift is in the sense expected if early
type galaxies at high redshift are slightly bluer than at the present epoch.
See the caption of Figure~\ref{fig:esel_cz} for further discussion.
}
\label{fig:esel_zz}
 \end{figure}

Figure \ref{fig:esel_zz} shows the correspondence between
spectroscopic and photometric redshift $z_E$ for galaxies
with $M_E < M_\star + 1$. It shows very good agreement, with
little scatter, though with a slight systematic offset at
$z \sim 0.5-1.0$ which we interpret as an evolutionary effect.

\subsection{Background (Source) Galaxy Sample}
\label{ssec:bg}

The background sample was selected to lie in a range of
significance $4 < \nu < 150$ (equivalent to limiting magnitudes of
$m_{I} \simeq 25$ and $m_{I} \simeq 21$ for a 
point source).
The resulting number of source galaxies was $147,933$.
The number of lens-source pairs in each redshift interval (summed
over all six pointings) is shown in 
Table~\ref{tab:data}.
In order to make accurate predictions for the shear variance
it is necessary to have an accurate model for the redshift
distribution for these faint galaxies or, more precisely, the
distribution of weight over redshift.   
The measurements used
here are not particularly deep, 
and there are nearly complete
redshift samples which probe the required magnitude range.  Here we
shall use the SSA22 field sample of Cowie which has the
greatest depth and spectroscopic completeness.

In both $I$- and $V$-band samples the weight is distributed over a range of several
magnitudes, with half of the weight attributed to galaxies 
brighter/fainter than $m_{I} \simeq 23.0$ and $m_{V} \simeq 24.2$.
The very faintest galaxies lie beyond the completion
limit of Cowie's sample, but the redshift distribution in a
band one magnitude wide about the median magnitude above is
well determined. To a first approximation, the
effect of variation of mean redshift with magnitude should 
cancel out, so we shall adopt the 
central band redshift distribution as appropriate for the full sample.
At this magnitude the samples are approximately 80\% complete,
and it is thought that the galaxies for which a redshift cannot
be obtained lie predominantly around $z = 1.5-2.0$.

\begin{figure}
\centering\epsfig{file=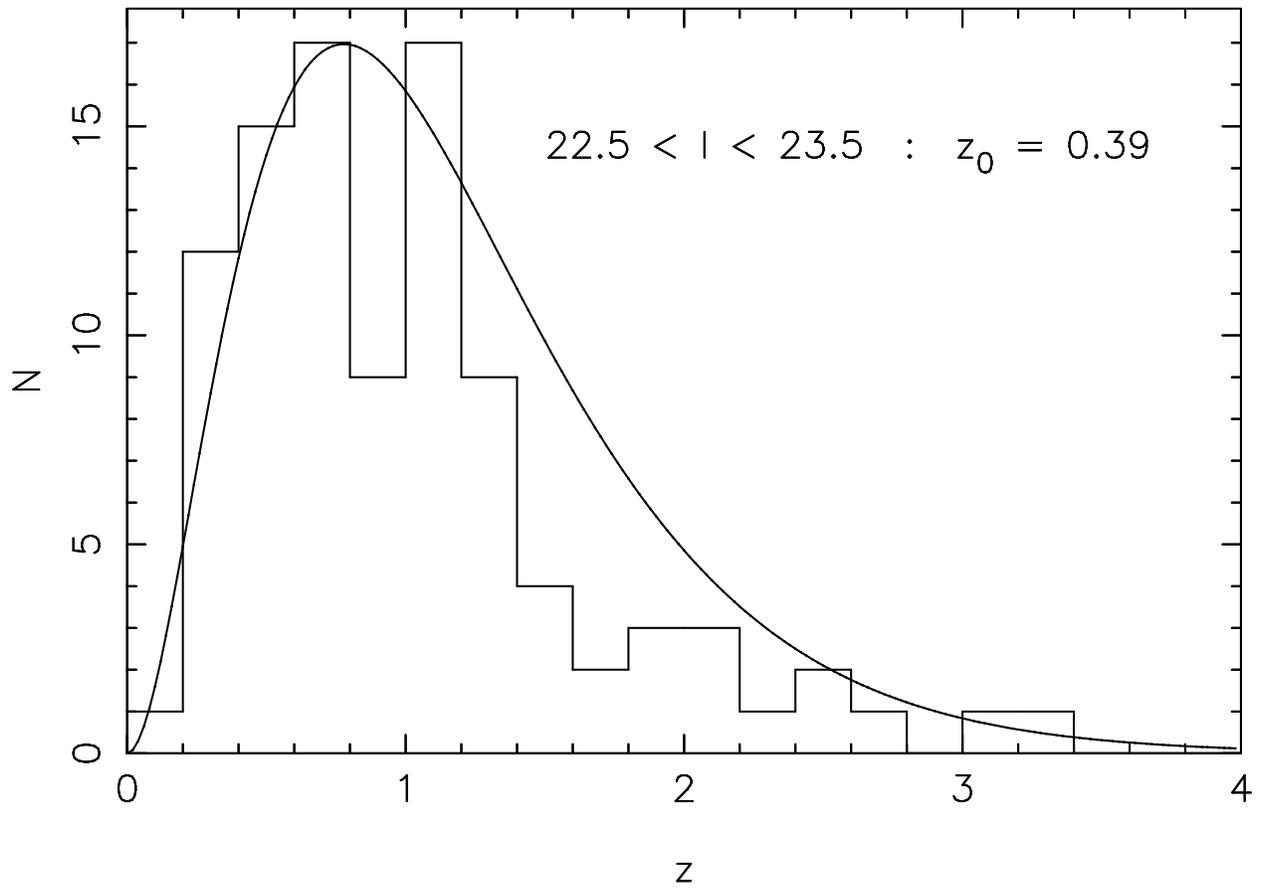,width=\figwidth}
\caption[nzmodel.I.tex]{The histogram shows the observed 
distribution of redshifts from Cowie.  The curve is
a model which allows for incompleteness by assuming
that the $\simeq 20\%$ of unmeasurable galaxies lie at
$z \simeq 1.8$.}
\label{fig:nz_I}
\end{figure}

We model the redshift distribution as
\begin{equation}
\label{eq:pz}
p(z)  = 0.5 z^{2} \exp(-z/\znought) / \znought^{3}
\end{equation}
for which the mean redshift is $\overline{z} = 3 z_0$ and the
median redshift is $z_{\rm median} = 2.67 z_0$.
This is also the analytic form used by \markcite{witt-00}{Wittman} {et~al.} (2000) and
others, and seems to adequately describe the data.
To allow for incompleteness we set the parameters $n_0$, $z_0$
of the model distribution to match the total number of
galaxies in the Cowie sample (with and without secure redshifts) and to match the
mean redshift with the unmeasurable objects assigned
a redshift $z = 1.8$.
Figure \ref{fig:nz_I} shows the redshift distribution 
for galaxies around $m_{I} = 23.0$
along with the incompleteness corrected model, which has
redshift scale parameter $z_0 = 0.39$.  The same calculation for
galaxies selected in a one magnitude wide band around $m_{V}=24.2$
yields a slightly smaller, though very similar, redshift
parameter $z_0 = 0.37$.  

\begin{figure}
\centering\epsfig{file=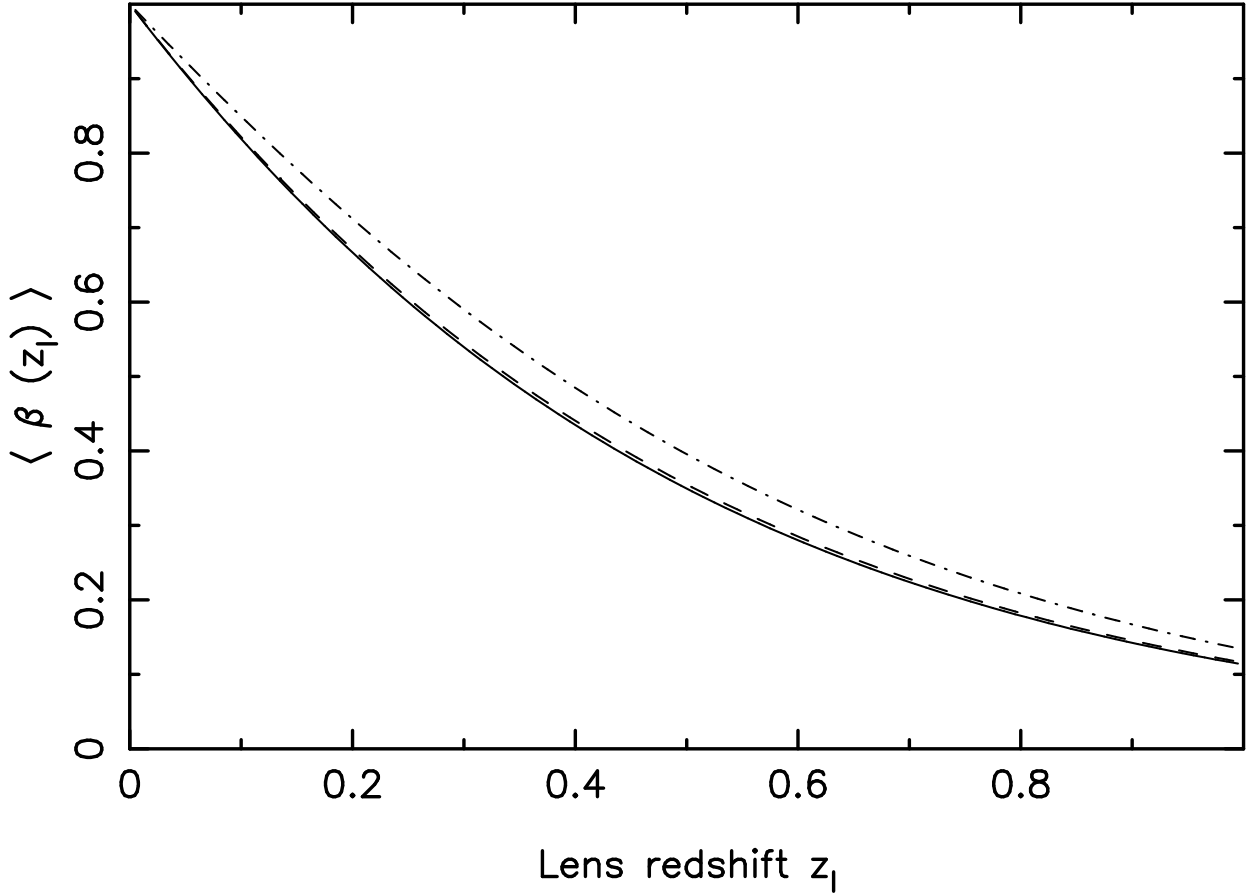,width=\figwidth}

\caption[plot_beta_varys_i23.ps]{
$\langle \beta (z_{l}) \rangle$ as a function of redshift and cosmology using the analytic approximation to an $m_{I} =23$ 
source galaxy redshift distribution (Figure~\ref{fig:nz_I}). Solid line is 
Einstein-deSitter ($\omegamnought = 1.0$, $\omegalnought = 0.0$), dotted is
open baryon ($\omegamnought = 0.05$, $\omegalnought = 0.0$), dot-dashed is flat lambda ($\omegamnought = 0.3$, $\omegalnought = 0.7$).
Non-lambda cosmologies have very similar $\langle \beta (z_{l}) \rangle$ values. Only the larger distances/volumes associated with lambda increase
$\langle \beta (z_{l}) \rangle$ significantly for any given redshift.
 \label{fig:beta}
}
\end{figure}

We now calculate $\langle \beta(z_{l}) \rangle$  as a function of lens redshift,
(see Table~\ref{tab:fitsFL} and Table~\ref{tab:fitsEdS} for the case of an
Einstein-de Sitter universe).
In Figure~\ref{fig:beta} we plot $\langle \beta \rangle$ as a function of lens
redshift for three cosmologies. The dashed line is flat lambda ($\omegamnought = 0.3, \omegalnought = 0.7$), the solid line is 
Einstein-deSitter ($\omegamnought = 1.0, \omegalnought = 0.0$), the dotted line is
open baryon ($\omegamnought = 0.05, \omegalnought = 0.0$).
Non-lambda cosmologies have very similar $\beta$ values. Only the larger distances to source galaxies associated with a cosmological
constant increase
$\beta$ significantly at any lens redshift. We return to the dependence of halo mass on cosmology
in \S\ref{sec:disc}.

\section{GALAXY DARK MATTER HALO MASSES}
\label{sec:results}

\subsection{Observed Tangential Shear Signal}
\label{ssec:gammatobs}

Having extracted a set of lens galaxies as described in \S~\ref{ssec:fg}
we now compute the tangential shear averaged over lens-source pairs.
However, not all lens galaxies will contribute
equally to the shear signal.  Insofar as galaxies have similar power-law
mass density profiles, the
massive lens galaxies cause more distortion
of source galaxies in proporton to their mass. Therefore, to optimize 
the signal to noise,
the shear contribution from each lens-source pair should be weighted 
by the mass of the lens.
At small radii the Faber-Jackson \markcite{fj-76}(Faber \& Jackson 1976, FJ) relation 
tells us that the mass 
at a given radius scales as $\sqrt{L}$. Later work has shown
that there is also an inter-dependence on a third parameter, 
the surface brightness of the galaxy, and that early types
describe a ``fundamental 
plane'' \markcite{dd-87,dress-87}({Djorgovski} \& {Davis} 1987; {Dressler} {et~al.} 1987) or ``fundamental band'' \markcite{glb-93}({Guzman}, {Lucey}, \& {Bower} 1993).
The FJ
correlation should be interpreted as a projection of this plane
onto the mass-luminosity plane.
However, the scatter introduced by neglecting surface
brightness and by assuming that $M\propto \sqrt{L}$
is slight compared to other uncertainties in the analysis.
Therefore, in the absence of information to the contrary we shall assume that $M \propto \sqrt{L}$ (\markcite{glb-93}{Guzman} {et~al.} conclude $M\propto L^{0.54}$) and we shall
also assume that this dependence continues to
larger radii. The weighted mean tangential shear is given by
\begin{equation}
\label{eq:meangammat}
\gamma_{T}(\theta) = {\sum w\gamma_{T} \over \sum w} = {\sum L^{1/2}\gamma_{T} \over \sum L^{1/2}}
\end{equation}
where the shear values have $W_{s}$ (equation~\ref{eq:gammaTdef}) incorporated.
\begin{figure}
\centering\epsfig{file=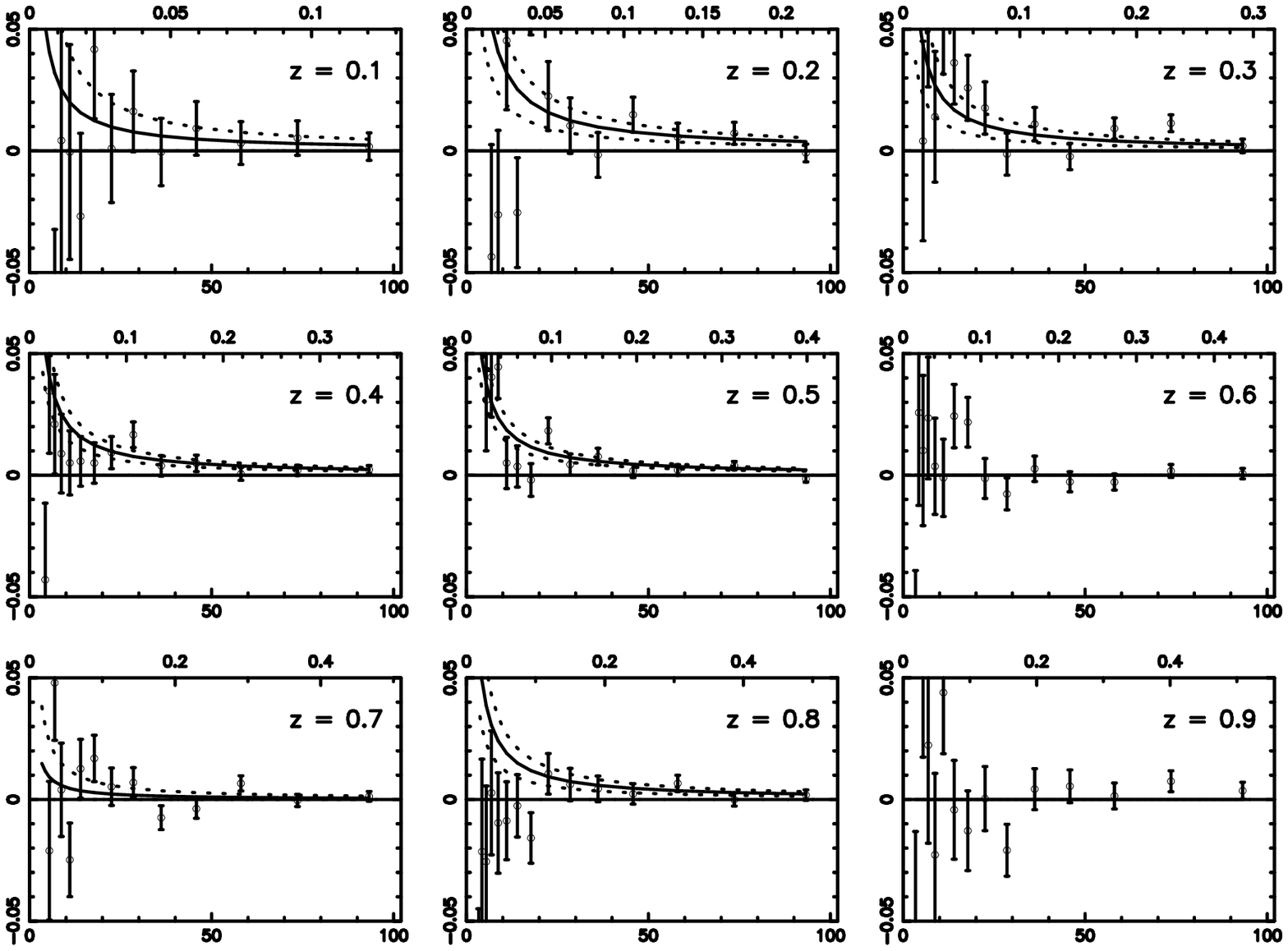,width=\figwidth}
\caption[galgal_betaeff_FL.ps]{
Mean tangential shear around early type (lens) galaxies. The foreground
galaxies have been color-selected and range in redshift from $z = 0.1$ to $z = 0.9$
in intervals of width $dz = 0.1$. The uncertainty has been  
calculated by rotating each source galaxy through 45 degrees. Lower axis 
shows the
lens-source galaxy projected radial separation in arcsec. Upper axis
shows the physical separation in $h^{-1}$ Mpc at the lens redshift 
assuming a flat lambda universe. The solid (dotted) line is the 
best ($\pm 1 \sigma$) fit to the data. 
 \label{fig:shearall}
}
\end{figure}

\begin{figure}
\centering\epsfig{file=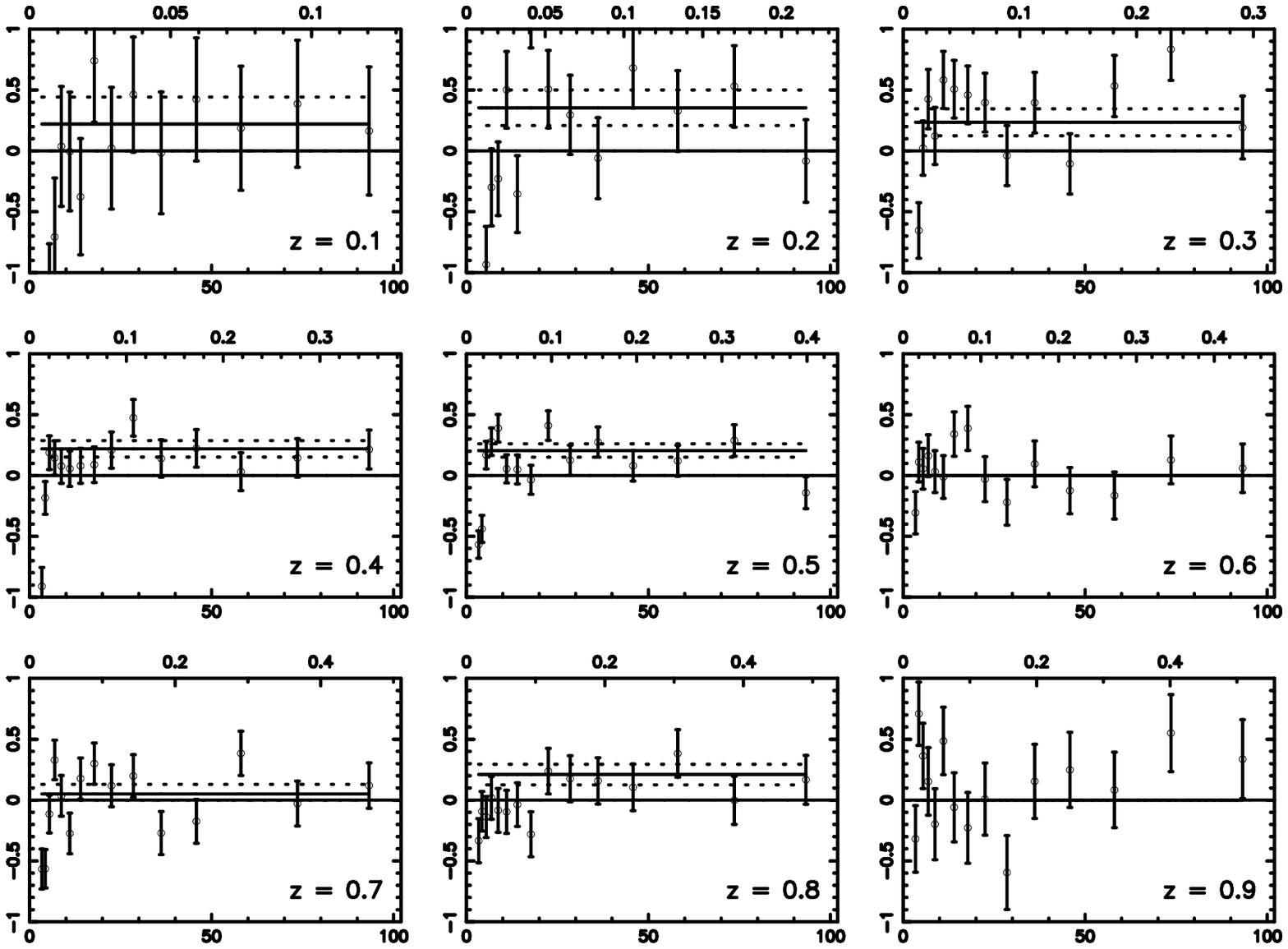,width=\figwidth}

\caption[galgal_betaeff_ettheta_FL.ps]{
Product of mean tangential shear around early type (lens) galaxies and
lens-source galaxy projected radial separation in arcsec. The foreground
galaxies have been color-selected and range in redshift from $z = 0.1$ to $z = 0.9$
in intervals of width $dz = 0.1$. The uncertainty has been  
calculated by rotating each source galaxy through 45 degrees. 
Lower axis shows the
lens-source galaxy projected radial separation in arcsec. Upper axis
shows the physical separation in $h^{-1}$ Mpc at the lens redshift 
assuming a flat lambda universe. The solid (dotted) line is the 
best ($\pm 1 \sigma$) fit to the data.
 \label{fig:shearthetaall}
}
\end{figure}

This weighted tangential shear is plotted in Figure~\ref{fig:shearall} 
for nine slices in lens redshift. 
The uncertainty (the variance in $\langle \gamma_{T} \rangle $) is 
calculated by rotating each source galaxy through 45 degrees.
The very highest and lowest redshift bins are rather noisy, but in general a
positive signal is seen.  As mentioned, for flat rotation curve halos the
shear falls as $1 / \theta$.  Figure \ref{fig:shearthetaall} 
shows the product $\theta \gamma_T$ which does indeed seem
to be roughly independent of radius.

The solid (dotted) line on Figures~\ref{fig:shearall} and~\ref{fig:shearthetaall}
shows the average $\theta \gamma_T$ ($\pm 1 \sigma$). The signal 
appears to be noisy and unreliable at small angular separation so 
we average points between $20\asec$ and $60\asec$. The best fit $\theta \gamma_T$ value is quoted (where positive)
for each redshift in Table~\ref{tab:fitsFL}. Also shown in the table
are the equivalent mean rotation velocity obtained using the 
$\langle \beta(z_l) \rangle$
values computed above and equation (\ref{eq:gammat}).

Since the $L_{\rm min}$ values vary with lens redshift because of the magnitude cut
discussed in section~\ref{ssec:fg}, one cannot compare
the different $\theta \gamma_T$ values directly (the lower redshift bins average over
somewhat fainter galaxies).  If however $M \propto L^{1/2}$ then the
mean $\theta \gamma_T$, 
and the equivalent rotation velocity, are equal to that for some
effective luminosity $L_{\rm eff}$.  This effective luminosity can be computed in two ways: either as a
direct sum over lens galaxies
\begin{equation}
\label{eq:LeffD}
\rtLeffrat_{\rm Direct} = \frac{\sum w \rtLstarrat} {\sum w} = \frac{\sum \Lstarrat} {\sum \rtLstarrat}
\end{equation}
or by integrating over the  luminosity function:
\begin{equation}
\label{eq:xsch}
\rtLeffrat_{\rm Schechter} =\frac{\int^{\infty}_{x_{\rm min}} x^{(1+\alpha)}\exp(-x) \,dx}{\int^{\infty}_{x_{\rm min}} x^{(0.5 + \alpha)} \exp(-x) \,dx }
\end{equation}
($x = \Lstarrat$), with parameters given by 2dF (Table~\ref{tab:2dF}).
These give very similar results (Table~\ref{tab:fitsFL}), so we use the
direct method henceforth.

Finally, given $v$ and $\rtLeffrat$ one can compute the equivalent mean rotation velocity for
an $\Lstar$ lens galaxy:
\begin{equation}
\label{eq:vstarlum}
\vstar^{2} = v^{2}/ \rtLeffrat
\end{equation}
This result is again strictly dependent on the assumption that masses
scale as $\sqrt{L}$, but given the limited range of absolute magnitudes
used here the result is only weakly dependent on this assumption. The last two columns of
Table~\ref{tab:fitsFL} show $\vnumstarratsq=\vnumratsq/\rtLeffrat$ and hence
$\vstar$ at each redshift.

\begin{figure}
\centering\epsfig{file=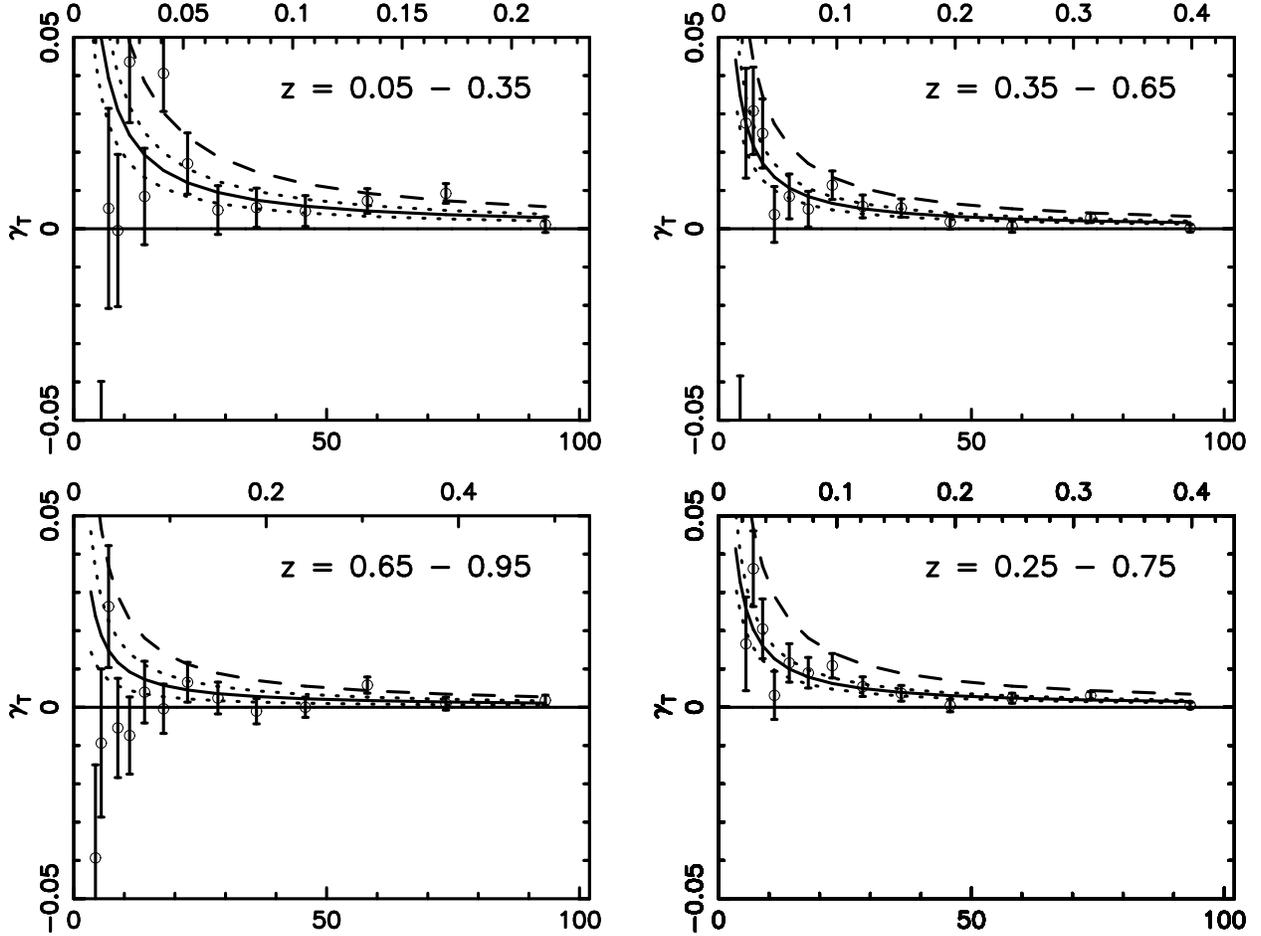,width=\figwidth}

\caption[galgal_betaeff_4_FL.ps]{
As for figure~\ref{fig:shearall} but for broader redshift range as indicated on each panel.
The solid (dotted) line is the 
best ($\pm 1 \sigma$) fit to the data. The dashed line is the predicted 
shear if small-scale values of rotation velocity from dynamical measurements 
are extrapolated to larger scales as discussed in \S\ref{ssec:dyn}. 
 \label{fig:shear4}
}

\end{figure}

\begin{figure}
\centering\epsfig{file=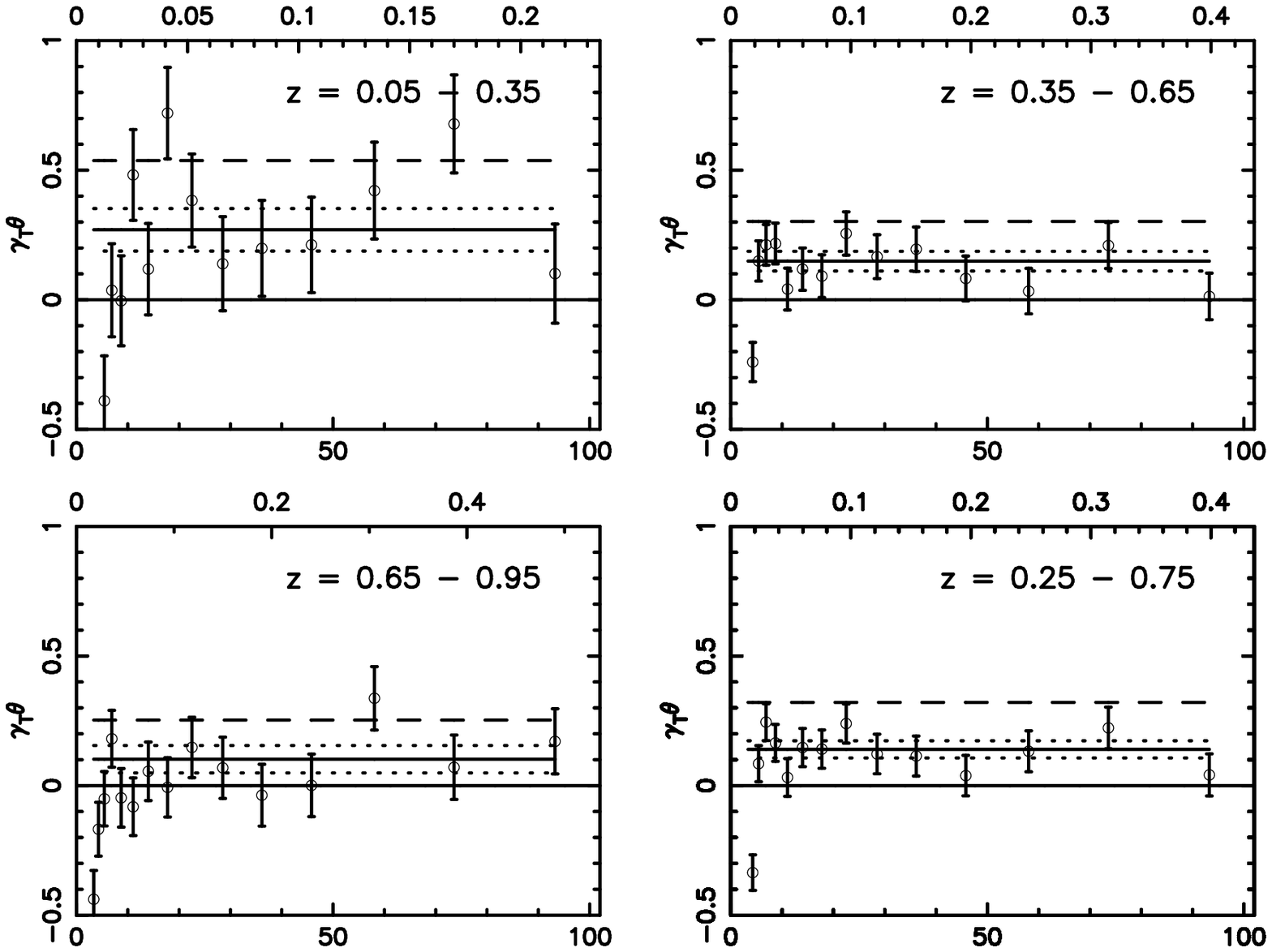,width=\figwidth}

\caption[galgal_betaeff_ettheta_4_FL.ps]{
As for Figure~\ref{fig:shearthetaall} but for broader redshift range as indicated on each panel. The solid (dotted) line is the 
best ($\pm 1 \sigma$) fit to the data.
The dashed line is the predicted 
shear if small-scale values of rotation velocity from dynamical measurements 
are extrapolated to larger scales as discussed in \S\ref{ssec:dyn}. 
 \label{fig:sheartheta4}
}
\end{figure}

While the narrow ($\Delta z = 0.1$) bins here give very good resolution in redshift,
the limited number of lens galaxies in each bin results in quite noisy
results.  To enhance the signal to noise, at the expense of a slight
loss in redshift resolution we now rebin the signal using 
coarser redshift bins.
Figures~\ref{fig:shear4} and~\ref{fig:sheartheta4} show the mean tangential shear signal,
$ \gamma_T $ and $ \theta \gamma_T $, using redshift bins of $dz = 0.3$ in first three panels.
Note that by by combining our data in this way we are summing 
over slightly different physical scales. In Table~\ref{tab:fitsFL} we again
calculate $\vnumstarratsq=\vnumratsq/\rtLeffrat$ and also $\vstar$.
We obtain values of $\vstar = 255^{+36}_{-42}$ for $z = 0.2\pm0.15$,
$\vstar = 253^{+30}_{-35}$ for $z = 0.5\pm0.15$ and  $\vstar = 228^{+53}_{-70}$ for $z = 0.8\pm0.15$.
Thus, it appears that there is little evolution in the mass of dark matter halos 
with redshift. In the final
panel of Figures~\ref{fig:shear4} and~\ref{fig:sheartheta4} we bin the signal for lens galaxies 
between $z = 0.25$ and $z = 0.75$. We conclude that $\vstar = 238^{+27}_{-30}$ for $z = 0.5\pm0.25$.

\begin{figure}
\centering\epsfig{file=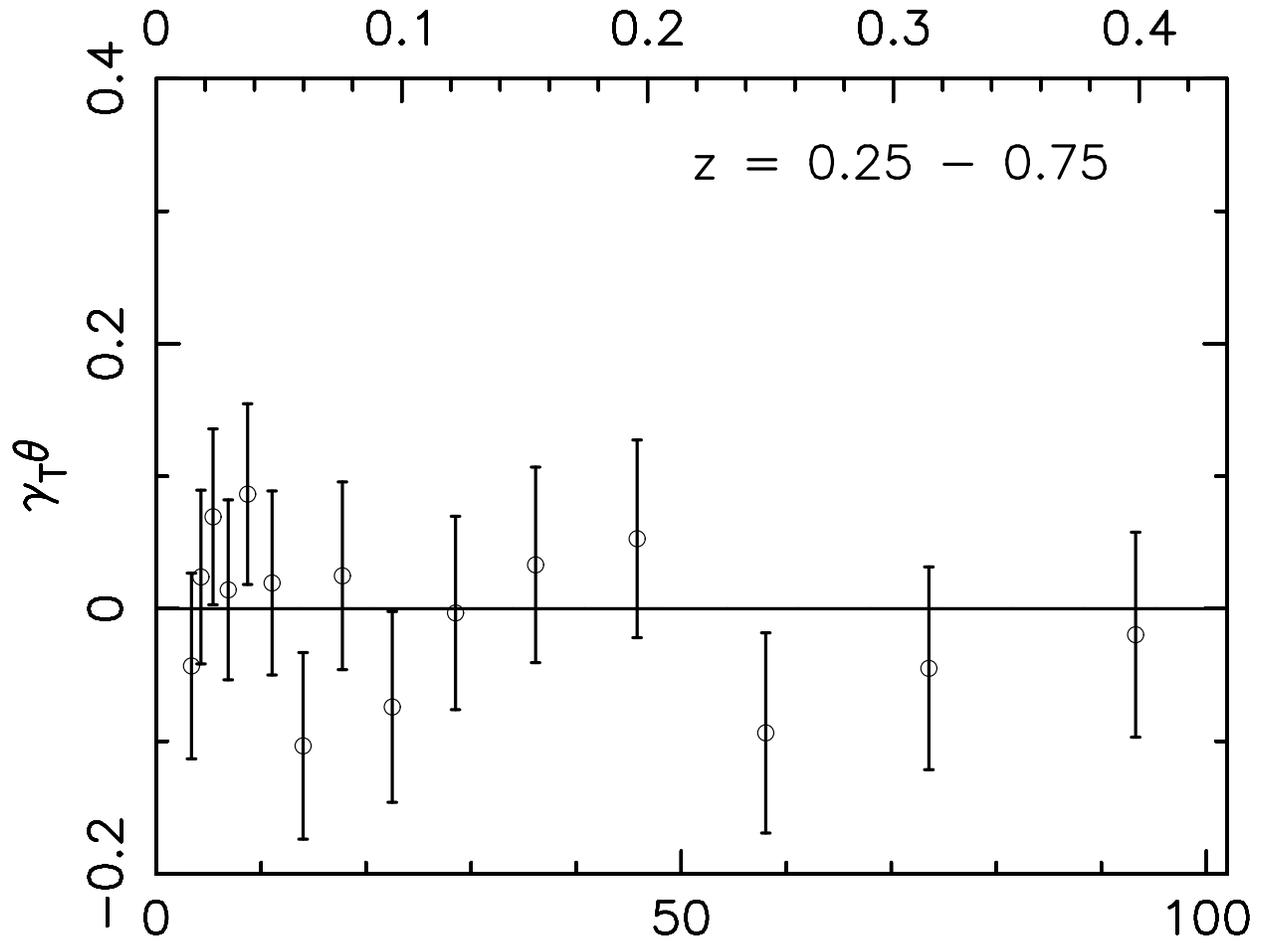,width=\figwidth}

\caption[galgal_betaeff_ettheta_null_1_FL.ps]{
As for Figure~\ref{fig:sheartheta4} but with galaxies rotated through 45 degrees.
The signal has disappeared as expected. 
 \label{fig:sheartheta1null}
}

\end{figure}

The signal strength that is being measured is small.
As a check for systematic errors, 
in Figure~\ref{fig:sheartheta1null} we use the same data as in Figure~\ref{fig:sheartheta4},
but rotate the galaxies through 45 degrees. As expected, there is no resultant signal 
causing us to conclude that systematic errors are negligible. 
(The interested reader is 
referred to \markcite{kwl-01}Kaiser, Wilson, \&  Luppino (2001a, Paper I) for a 
description of our careful corrections for systematics.)

\clearpage

\begin{deluxetable}{cccccccl}
\tablewidth{0pt}
\tabletypesize{\scriptsize}
\rotate
\tablecaption{Model Parameters for a Flat Lambda ($\omegamnought = 0.3, \omegalnought = 0.7$) Cosmology
\label{tab:fitsFL}
}
\tablewidth{0pt}
\tablehead{
\colhead{Lens Redshift}  & 
\colhead{$<\gamma_{T}\theta>(20-60'')$} &
\colhead{$\langle \beta (z_{l}) \rangle$}  &
\colhead{$\vnumratsq$} &
\colhead{$\rtLeffrat_{\rm Direct}$} &
\colhead{$\rtLeffrat_{\rm Schechter}$} &
\colhead{$\vnumstarratsq$}&
\colhead{$\vstar $}
}
\startdata
$0.1\pm0.05$  & $0.220\pm0.222$ & $0.849$ & $0.278\pm0.278$ & $0.487$ & $0.675$ & $0.570\pm0.570$ & $272^{+113}_{-272}$ 
\\
$0.2\pm0.05$  & $0.354\pm0.146$ & $0.711$ & $0.533\pm0.220$ & $0.887$ & $0.872$ & $0.601\pm0.248$ & $279^{+53}_{-65}$ 
\\
$0.3\pm0.05$  & $0.235\pm0.111$ & $0.589$ & $0.427\pm0.202$ & $0.905$ & $1.022$ & $0.472\pm0.223$ & $247^{+53}_{-68}$ 
\\
$0.4\pm0.05$  & $0.219\pm0.068$ & $0.484$ & $0.485\pm0.150$ & $0.804$ & $0.872$ & $0.603\pm0.187$ & $279^{+40}_{-47}$ 
\\
$0.5\pm0.05$  & $0.205\pm0.055$ & $0.395$ & $0.556\pm0.149$ & $0.743$ & $0.615$ & $0.748\pm0.201$ & $311^{+39}_{-45}$ 
\\
$0.6\pm0.05$  & $\ldots$        & $0.320$ & $\ldots$        & $0.892$ & $0.702$ & $\ldots$         & $\ldots$ 
\\
$0.7\pm0.05$  & $0.050\pm0.079$ & $0.259$ & $0.207\pm0.327$ & $1.065$ & $0.808$ & $0.194\pm0.307$  & $159^{+96}_{-159}$ 
\\
$0.8\pm0.05$  & $0.210\pm0.085$ & $0.208$ & $1.082\pm0.438$ & $1.244$ & $0.946$ & $0.870\pm0.352$  & $336^{+62}_{-77}$ 
\\
$0.9\pm0.05$  & $\ldots$        & $0.167$ & $\ldots$        & $1.419$ & $1.127$ & $\ldots$         & $\ldots$ 
\\
\\
$0.2\pm0.15$  & $0.270\pm0.082$ & $0.664$ & $0.435\pm0.132$ & $0.866$ & $\ldots$ & $0.503\pm0.153$ & $255^{+36}_{-42}$ 
\\
$0.5\pm0.15$  & $0.149\pm0.038$ & $0.406$ & $0.393\pm0.100$ & $0.797$ & $\ldots$ & $0.493\pm0.126$ & $253^{+30}_{-35}$ 
\\
$0.8\pm0.15$  & $0.102\pm0.053$ & $0.225$ & $0.485\pm0.252$ & $1.205$ & $\ldots$ & $0.403\pm0.209$ & $228^{+53}_{-70}$ 
\\
\\
$0.5\pm0.25$  & $0.140\pm0.033$ & $0.398$ & $0.377\pm0.089$ & $0.864$ & $\ldots$ & $0.436\pm0.103$ & $238^{+27}_{-30}$ 
\\
\enddata
\end{deluxetable}

\clearpage

\begin{deluxetable}{cccccccl}
\tablewidth{0pt}
\tabletypesize{\scriptsize}
\rotate
\tablecaption{Model Parameters for an Einstein-de Sitter ($\omegamnought = 1.0, \omegalnought = 0.0$) Cosmology
\label{tab:fitsEdS}
}
\tablewidth{0pt}
\tablehead{
\colhead{Lens Redshift}  & 
\colhead{$<\gamma_{T}\theta>(20-60'')$} &
\colhead{$\langle \beta (z_{l}) \rangle$}  &
\colhead{$\vnumratsq$} &
\colhead{$\rtLeffrat_{\rm Direct}$} &
\colhead{$\rtLeffrat_{\rm Schechter}$} &
\colhead{$\vnumstarratsq$}&
\colhead{$\vstar $}
}
\startdata
$0.1\pm0.05$  & $0.206\pm0.233$ & $0.819$ & $0.269\pm0.305$ & $0.475$ & $0.676$ & $0.567\pm0.641$ & $271^{+125}_{-271}$ 
\\
$0.2\pm0.05$  & $0.374\pm0.158$ & $0.667$ & $0.601\pm0.254$ & $0.848$ & $0.872$ & $0.708\pm0.299$ & $303^{+58}_{-73}$ 
\\
$0.3\pm0.05$  & $0.241\pm0.133$ & $0.540$ & $0.478\pm0.264$ & $0.862$ & $1.022$ & $0.555\pm0.306$ & $268^{+66}_{-89}$ 
\\
$0.4\pm0.05$  & $0.215\pm0.080$ & $0.435$ & $0.529\pm0.197$ & $0.745$ & $0.872$ & $0.710\pm0.264$ & $303^{+52}_{-63}$ 
\\
$0.5\pm0.05$  & $0.205\pm0.055$ & $0.350$ & $0.628\pm0.168$ & $0.618$ & $0.567$ & $1.016\pm0.273$ & $363^{+46}_{-52}$ 
\\
$0.6\pm0.05$  & $\ldots$        & $0.280$ & $\ldots$        & $0.725$ & $0.635$ & $\ldots$         & $\ldots$ 
\\
$0.7\pm0.05$  & $0.050\pm0.079$ & $0.224$ & $0.239\pm0.378$ & $0.849$ & $0.712$ & $0.282\pm0.445$  & $191^{+116}_{-191}$ 
\\
$0.8\pm0.05$  & $0.210\pm0.085$ & $0.179$ & $1.258\pm0.509$ & $0.973$ & $0.808$ & $1.293\pm0.523$  & $409^{+76}_{-94}$ 
\\
$0.9\pm0.05$  & $\ldots$        & $0.142$ & $\ldots$        & $1.092$ & $0.929$ & $\ldots$         & $\ldots$ 
\\
\\
$0.2\pm0.15$  & $0.282\pm0.093$ & $0.630$ & $0.479\pm0.158$ & $0.820$ & $\ldots$ & $0.584\pm0.193$ & $275^{+42}_{-50}$ 
\\
$0.5\pm0.15$  & $0.141\pm0.040$ & $0.354$ & $0.426\pm0.121$ & $0.680$ & $\ldots$ & $0.627\pm0.178$ & $285^{+38}_{-44}$ 
\\
$0.8\pm0.15$  & $0.102\pm0.053$ & $0.194$ & $0.563\pm0.293$ & $0.945$ & $\ldots$ & $0.596\pm0.310$ & $278^{+65}_{-85}$ 
\\
\\
$0.5\pm0.25$  & $0.131\pm0.035$ & $0.343$ & $0.409\pm0.109$ & $0.732$ & $\ldots$ & $0.559\pm0.149$ & $269^{+34}_{-39}$ 
\\
\enddata
\end{deluxetable}

\clearpage


\section{DISCUSSION}
\label{sec:disc}

In the previous sections we have shown that with our $I$ and $V$-band
CFHT data we can select a sample of bright early type galaxies,
determine their redshifts to a reasonable degree of precision, and
measure the shear that they produce in faint background galaxies over quite
a range of angular scales and lens redshifts.  We find little evolution of the
halos with redshift. We also find that the radial dependence of the shear
is consistent with roughly flat rotation curve halos.   Our results
imply mean rotation velocities for $\Lstar$ galaxy halos at
$r \sim 50-200 h^{-1}$kpc of $\vstar = 238^{+27}_{-30}\kms$.  This number is
dependent on the assumption that the mass at a given radius
scales as the square root of the luminosity, as is known to be
the case at much smaller scales from the Faber-Jackson  relation.
However, since our lens galaxy sample
is restricted to relatively bright galaxies --- within a magnitude
or so of $L_\star$ --- we expect this dependence to be rather weak.
We now discuss some of the implications of this result. We compute the
mass-to-light ratio and the contribution to the total density
from these halos.  We compare our results with dynamical measurements at
smaller scales, with X-ray and other lensing measurements at similar
scales to these we can reliably measure.  We also
compare the properties of these halos to those found in 
numerical cosmological simulations. Finally, we discuss  
uncertainties due to evolution and cosmology.

\subsection{$M/L$ and Contribution to $\Omega_0$}

An $\Lstar$ galaxy halo with $\vstar = 238$ contains 
$1.31\times10^{12}(r/100\hkpc)h^{-1}\Mo$ within a 
radius of $r$, since $M(r) = \vstarsq r/G$. An $\Lstar$ galaxy has a luminosity of 
$1.09\times10^{10}h^{-2}\Lo^{B}$, so
the mass to light ratio  is $\mlb = 121\pm28h(r/100\hkpc)$, or
about $\mlb \sim 250 h$ at the outermost points we can reliably measure.

We can compute the contribution of these halos to the total density
of the Universe. This is, of course, only a partial contribution since
only early type galaxies are counted --- though they may well in fact
account for the majority of the mass --- and because here we have deliberately
restricted attention to relatively small scales $\lesssim 200 h^{-1}$kpc.   
We shall assume, as above, that $M \propto \sqrt{L}$,
so
\begin{equation}
M(r) = M_\star(r) \sqrt{L / L_\star},
\end{equation}
where $M_\star(r)$ is the mass profile for an $L_\star$ galaxy, and
the density is then
\begin{equation}
\rho = M_\star(r) \int dL \; \phi_E(L) \sqrt{L / L_\star}
= M_\star(r) \phi_{E\star} \Gamma(\alpha + 3/2).
\end{equation}
With the numbers from Table~\ref{tab:2dF} we find 
this constitutes $\Omega = 0.04\pm0.01(r/100\hkpc)$ of closure density.

\subsection{Comparison with Small Scale Dynamics}
\label{ssec:dyn}
Stellar velocity dispersions in ellipticals probe the mass on scales of
a few kpc --- much smaller than the scales we are measuring --- and
yield the Faber-Jackson relationship (that $L \propto \sigma_{v}^{4}$ \markcite{fj-76}(Faber \& Jackson 1976)).
A distillation of these and later studies by \markcite{ft-91}Fukugita \& Turner (1991)
gives $\sigma_{v\star} = 210\kms$ 
(the mean of their E/S0 line of sight 
stellar velocity dispersions).
If we assume that the stars
are test particles on finite orbits in a roughly flat rotation curve
halo, the $L_\star$ rotation velocity at a scale of a few kpc
is then $\sqrt{3} \times 210 \simeq 360\kms$.  The corresponding mass is larger
by a factor of $2-3$ than the value we measure on scales of $50-200\hkpc$.
For interest, in Figures~\ref{fig:shear4} and~\ref{fig:sheartheta4} we plot 
(dotted line) the signal which would be obtained from a
galaxy with the same effective luminosity $\rtLeffrat$ (Table~\ref{tab:fitsFL})
as our galaxies, but with the rotation velocity of $360\kms$ (the value
determined on
small-scales from Faber-Jackson measurements).
Clearly, in all cases, the predicted signal is larger than the measured signal.
Thus, while both small and large-scale measurements are individually consistent with flat
rotation curves, if we combine them they 
suggest that the mean density profiles are actually falling off slightly faster than
$\rho \propto r^{-2}$.  If we say that our measurements are probing radii
a factor $\sim 30$ larger than the stellar dynamical measurements, and
that our $v^2$ is about 2.3 times smaller, then the mean profile over
this range is $\rho \propto r^{-(2 + \epsilon)}$ with
$\epsilon = \ln(2.3) / \ln(30) \simeq 0.24$, so $\rho \propto r^{-2.2}$.
Note that such a small departure from a pure flat rotation curve would be 
impossible to detect from either set of measurements alone.  Neither does
the small departure from $\rho \propto r^{-2}$ seriously
invalidate e.g.~equation (\ref{eq:gammat}).

\begin{figure}
\centering\epsfig{file=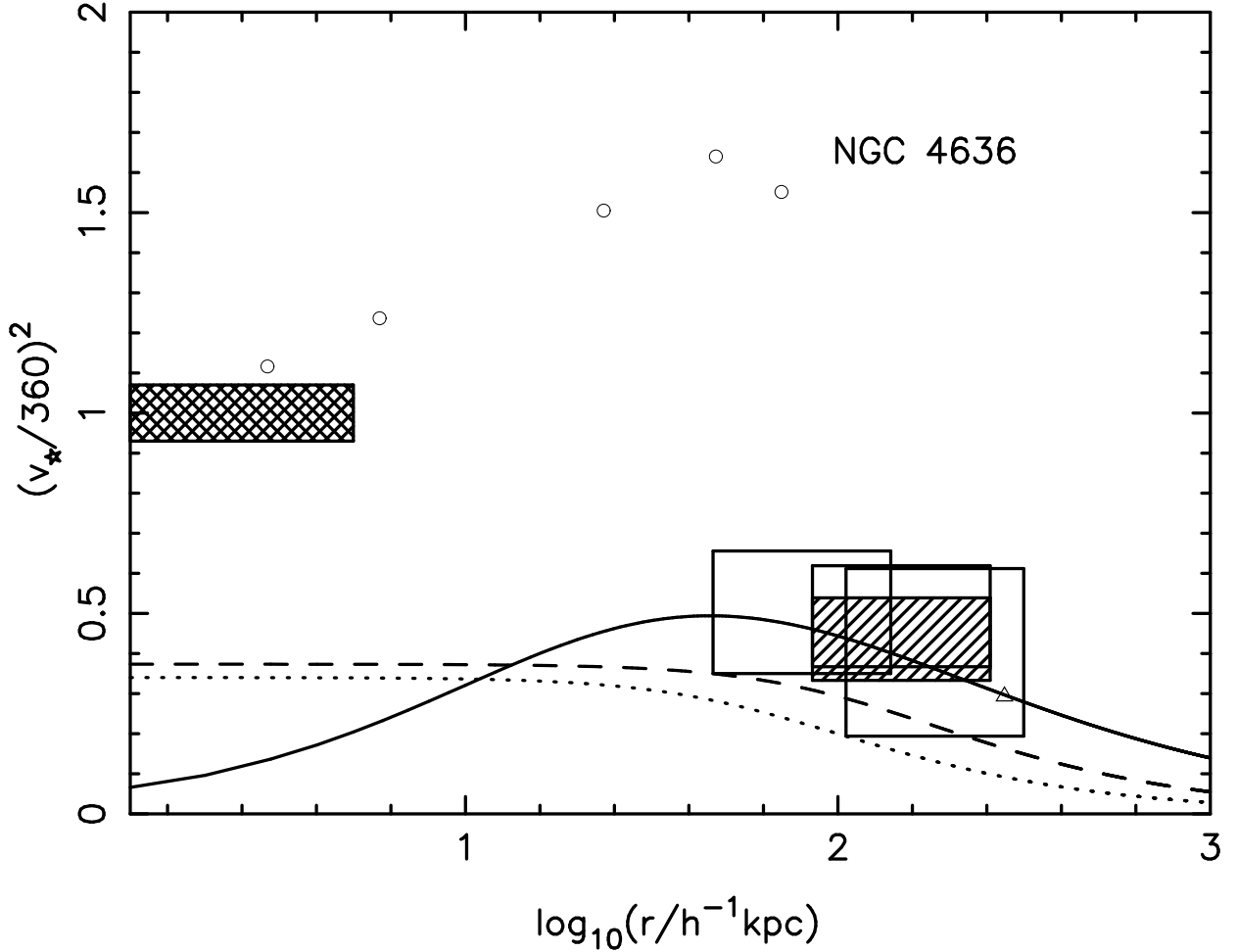,width=\figwidth}
\caption[mr.ps]{
Square of rotation velocity as a function of radius. The hashed rectangle is the value 
determined for the central region of an $\Lstar$ galaxy~\markcite{ft-91}(Fukugita \& Turner 1991). 
The three outlined rectangles are the values obtained from this work
(Table~\ref{tab:fitsFL}) for an $\Lstar$ galaxy at redshifts $0.2, 0.5, 0.8\pm0.15$. The 
striped rectangle is the value from this work
(Table~\ref{tab:fitsFL}) for an $\Lstar$ galaxy at redshift $ 0.5\pm0.25$. The
 circles are the values determined
by \markcite{mush-94}{Mushotzky} {et~al.} (1994) from X-ray measurements of NGC 4636 (assuming $\Hnought = 70\kmsMpc$).
The triangle is the value of halo mass from \markcite{jenk-01}{Jenkins} {et~al.} (2001)
with the same abundance as an $\Lstar$ galaxy.
The solid line is the rotation curve for an NFW profile \markcite{nfw-97}({Navarro}, {Frenk}, \& {White} 1997) 
 with normalization parameter chosen to intercept the triangle. The dashed and dotted
lines are the best fits from \markcite{brain-96}Brainerd {et~al.} (1996) and \markcite{hud-98}Hudson {et~al.} (1998) 
to a parametric model for {\em late} types.
 \label{fig:mr}
}
\end{figure}

In Figure~\ref{fig:mr} we plot various 
rotation velocity estimates as a function of radius.
The hashed rectangle at the smallest scale on Figure~\ref{fig:mr} is from~\markcite{ft-91}Fukugita \& Turner.
The three outlined rectangles are the values from this work
(Table~\ref{tab:fitsFL}) for an $\Lstar$ galaxy at redshifts $0.2, 0.5, 0.8\pm0.15$. The 
striped rectangle is the values from this work
(Table~\ref{tab:fitsFL}) for an $\Lstar$ galaxy at redshift $0.5\pm0.25$. 

\subsection{Comparison with X-ray Halos}

As discussed in the Introduction, elliptical galaxy halo masses have been
determined in a small number of cases from X-ray observations. 
In the best studied case of NGC 4636 \markcite{mush-94}({Mushotzky} {et~al.} 1994) the mass is very large and grows with radius
faster than $M \propto r$ out to $r \sim 100$kpc. These results
are shown as circles in Figure \ref{fig:mr} (assuming $\Hnought = 70\kmsMpc$).
If representative of
elliptical galaxies in general this would have weighty implications \markcite{bld-95}({Bahcall}, {Lubin}, \& {Dorman} 1995).
However, there is clearly some question as to whether these galaxies,
which have abnormally bright and extended X-ray emission are typical.
Our results strongly suggest that NGC 4636 is indeed an atypical object, being far more
massive than typical elliptical galaxies, and consequently that the
contribution of these galaxies to the \markcite{bld-95}{Bahcall} {et~al.} accounting needs to be revised
substantially downwards by a factor $2-3$.

\subsection{Comparison with Cosmological Simulations}

We now compare the properties of elliptical galaxy halos
with halos of the same abundance in numerical cosmological
simulations.  This should be a valid test of theoretical models since we
have measured the mass on scales that should be accurately
modeled on the computer, and should be little affected by
gas dynamics and star formation.  

With the 2dF luminosity function parameters the
number density of ellipticals brighter than $L_\star$ is
\begin{equation}
N(> L_*) = \phi_\star \int\limits_1^\infty dy \; y^{\alpha} e^{-y} 
\simeq 0.25 \times \phi_\star \simeq 2.2 \times 10^{-3} h^3 {\rm Mpc}^{-3}
\end{equation}

The differential mass function from the high resolution
$\Lambda$CDM simulations of \markcite{jenk-01}{Jenkins} {et~al.} (2001, their figure 2) probe the
relevant mass scales.  Integrating these to obtain
the cumulative mass function we find that $N(>M) = N(> L_*)$ for
$M \simeq 2.44 \times 10^{12} h^{-1} M_\odot$.  
The mass here is the mass for an overdensity of $324$, which,
with $\Omega_{{\rm m}0} = 0.3$, corresponds to radius of $r \simeq 279 h^{-1}$kpc,
and to a rotation velocity at that radius of $v \simeq 194\kms$
(shown by the triangle in Figure~\ref{fig:mr}).
\markcite{nfw-97}{Navarro} {et~al.} (1997) find that halos in this mass range in their $\Lambda$CDM
simulations are well described by their universal model with
concentration parameter $c \simeq 10$ (their figure 6), and to match
\markcite{jenk-01}{Jenkins} {et~al.}
rotation velocity requires $v_{200} \simeq 210\kms$.
The rotation curve profile for such a model is shown as
the solid line in  Figure~\ref{fig:mr} and matches our measured values extremely well .

This comparison --- simply matching the cumulative number density 
of halos to that of $L>L_\star$ elliptical ---
should not be considered definitive, but it is the best one can
do with the published numerical results.  This comparison could
be improved by using
semi-analytic galaxy formation
to identify plausible candidates for elliptical galaxies and
then computing the average mass profile around these.  

\subsection{Comparison with other Galaxy-Galaxy Lensing Studies}

A number of other groups have measured $\vstar$ from galaxy-galaxy lensing. 
Early studies had low signal-to-noise and results were typically presented as constraints on parameterized models.
Values of
$\vstar = 220\pm80\kms$ \markcite{brain-96}(Brainerd {et~al.} 1996),
$\vstar = 262^{+42}_{-49}\kms$ \markcite{datys-96}(Dell'Antonio \& Tyson 1996), and
$\vstar = 210\pm40\kms$ \markcite{hud-98}(Hudson {et~al.} 1998) were obtained. 
\markcite{datys-96}Dell'Antonio \& Tyson measured a signal on very
small scales (within 
a projected radius of $5\asec$) so their value is not directly comparable to that obtained here.  
\markcite{brain-96}Brainerd {et~al.} and \markcite{hud-98}Hudson {et~al.} both used the same parametric model for halo mass 
(equation 3.4 of  \markcite{brain-96}Brainerd {et~al.}).
In Figure~\ref{fig:mr}, we plot rotation velocity with radius using the
 best fit solution from \markcite{brain-96}Brainerd {et~al.} (dashed line)
and of \markcite{hud-98}Hudson {et~al.} (dotted line).
It should be noted that \markcite{brain-96}Brainerd {et~al.} (1996), \markcite{datys-96}Dell'Antonio \& Tyson (1996), and \markcite{hud-98}Hudson {et~al.} (1998) were all measuring halo rotation velocities for
primarily late type galaxies and one would expect lower values for their halo masses than those obtained from our measurements for early types. (\markcite{ft-91}Fukugita \& Turner (1991) find $\vstar = 204\kms$ for late type galaxies from the Tully-Fisher \markcite{tf-77}({Tully} \& {Fisher} 1977) relationship). 

\markcite{fisch-00}{Fischer} {et~al.} (2000) were the first group to obtain sufficiently high signal-to-noise
to fit a power-law directly to their data. Using preliminary data from the
Sloan Digital Sky Survey they obtained a signal out to several hundred 
arcsec and measured a  rotation velocity of $\vstar = 240\pm28\kms$.
(Note that the uncertainty here is a $95\%$ 
confidence limit not a $1\sigma$ uncertainty as for the other groups.) 
The interpretation of their results is somewhat 
complex because their lens sample is a mixture of early and late types, but the
power and potential of galaxy-galaxy lensing was convincingly demonstrated.

\subsection{Uncertainties due to Evolution and Cosmology} 

The analysis in this paper assumed that $\Lstar$ does not evolve with redshift. Based on our knowledge about early type galaxy evolution with redshift 
this does not seem a grossly inaccurate assumption. From the Canada-France
redshift survey, \markcite{lilly-95}{Lilly} {et~al.} (1995) found that their red (redder than 
present-day Sbc and hence early type) 
sample was consistent with no change in $\Lstar$ 
between $z\sim0.8$ and $z\sim0.3$ (their red sample was also consistent
with a change of \emph{at most} a few tenths of a magnitude)

We note that a brightening of 0.5 magnitude
in $\Lstar$ for the highest redshift sample ($z = 0.8\pm0.15$) 
(which might be feasible due to passive evolution) 
would induce a small ($\sim25\%$) increase in $\vstar^{2}$.

We assumed a flat lambda cosmology. If, for example we had assumed an
Einstein-de Sitter cosmology (Table~\ref{tab:fitsEdS}) in preference
to flat lambda, the inferred
values of $\vstar$ would still be approximately constant with redshift but would increase to $\vstar = 275^{+42}_{-50}\kms$ 
for $z = 0.2\pm0.15$,
$\vstar = 285^{+38}_{-44}\kms$ for $z = 0.5\pm0.15$ and  $\vstar = 278^{+65}_{-85}\kms$ for $z = 0.8\pm0.15$.
The increase in $\vstar$ in such a universe is primarily caused by smaller $\langle \beta \rangle$ values in this cosmology
(Table~\ref{tab:fitsEdS} and Figure~\ref{fig:beta}). We would conclude a rotation velocity of
$\vstar = 269^{+34}_{-39}\kms$ for $z = 0.5\pm0.25$ for this cosmology.

\section{CONCLUSIONS}
\label{sec:conc}

Unlike previous galaxy-galaxy lensing analyses we showed that it was possible to use
colors and magnitudes to cleanly select one type of lens galaxy (in this case
bright early types). By measuring a weighted mean tangential shear which 
decreased roughly as $1/\theta$ we concluded that early type galaxies have 
approximately flat rotation curve halos extending out to several hundred $\hkpc$. 
By assuming an $M \propto L^{1/2}$ relationship we inferred a rotation velocity
for an $\Lstar$ galaxy of 
$v_\star =  238^{+27}_{-30}\kms$ for a flat lambda 
($\omeganought = 0.3, \lambdanought = 0.7$) cosmology
($v_\star = 269^{+34}_{-39}\kms$ for Einstein-de Sitter)  with 
little evidence for evolution with redshift. 
These halo masses are somewhat ($2-3$ times)
lower than a simple perfectly flat rotation curve extrapolation 
from small-scale dynamical measurements.  They are also
considerably  lower than the masses of halos found from the
best studied X-ray halos although we note that the best X-ray example is
likely an atypical object. Interestingly, the 
values of halo mass determined
from galaxy-galaxy lensing and and the masses of halos of the same
abundance in lambda-CDM simulations agree remarkably well. We note, however,
that for an optimum comparison, halo masses should be determined as a 
function of redshift directly from the simulations. 
 
Finally, we determined a mass-to-light ratio for galaxy halos of 
$\mlb = 121\pm28h(r/100\hkpc)$ (for $\Lstar$ galaxies) and found 
that these halos constitute 
$\Omega \simeq 0.04 \pm 0.01(r/100\hkpc)$ of closure density.

In the foreseeable future it will be possible to measure
early type galaxy halo masses rather more precisely.
The color-redshift degeneracy (illustrated by Figure~\ref{fig:lf})
could be broken by the availability of a larger number of 
passbands to provide 
photometric redshifts such as will be provided by the Hawaii Lensing Survey, the Deep Lens Survey, or 
the Megacam/Terapix consortium. 
More 
preferable would be spectroscopic redshift determinations such as will be 
provided by the Sloan Digital Sky Survey.
The greater range of absolute
luminosity then available (limited here to $M \sim M_* \pm 1$) will allow
mass-to-luminosity dependence 
(assumed in this work to be $M \propto \sqrt{L}$) 
to be determined more precisely. 
Moreover, increased numbers of early type lens galaxies will reduce 
uncertainties in the measurement of tangential shear and allow any variation
in a $1 /\theta$ (\ie\ flat) rotation curve galaxy halo profile 
to be determined. Finally, the availability of $>2$-passband data
will also allow photometric redshifts for late type galaxies and a similar 
investigation to be undertaken into the properties of their halos.

\acknowledgements GW gratefully acknowledges financial support from the
estate of Beatrice Watson Parrent and from Mr. \& Mrs. Frank W. Hustace, Jr. whilst Parrent Fellow at UH. This work was supported by NSF grant
AST99-70805.


\clearpage


\begin{thebibliography}{}

\bibitem[{Bahcall} \& {Tremaine} 1981]{bt-81}
{Bahcall}, J.~N. \& {Tremaine}, S. 1981, \apj, 244, 805

\bibitem[{Bahcall}, {Lubin}, \& {Dorman} 1995]{bld-95}
{Bahcall}, N.~A., {Lubin}, L.~M., \& {Dorman}, V. 1995, \apjl, 447, L81

\bibitem[{Bosma} 1981]{bos-81}
{Bosma}, A. 1981, \aj, 86, 1825

\bibitem[Brainerd, Blandford, \& Smail 1996]{brain-96}
Brainerd, T.~G., Blandford, R.~D., \& Smail, I. 1996, ApJ, 466, 623

\bibitem[Coleman, Wu, \& Weedman 1980]{cole-80}
Coleman, G.~D., Wu, C., \& Weedman, D.~W. 1980, ApJS., 43, 393

\bibitem[Cowie, Gardner, Hu, Songaila, Hodapp, \&  Wainscoat 1994]{cghskw-94}
Cowie, L.~L., Gardner, J.~P., Hu, E.~M., Songaila, A., Hodapp, K.~W., \&  Wainscoat, R.~J. 1994, ApJ, 434, 114

\bibitem[Cowie, Songaila, \& Barger 1999]{csb-99}
Cowie, L.~L., Songaila, A., \& Barger, A.~J. 1999, AJ, 118, 603 (CSB)

\bibitem[{Cowie}, {Songaila}, {Hu}, \&  {Cohen} 1996]{cshc-96}
{Cowie}, L.~L., {Songaila}, A., {Hu}, E.~M., \& {Cohen}, J.~G. 1996, \aj, 112,  839

\bibitem[{Davis} \& {Peebles} 1983]{dp-83}
{Davis}, M. \& {Peebles}, P. J.~E. 1983, \apj, 267, 465

\bibitem[Dell'Antonio \& Tyson 1996]{datys-96}
Dell'Antonio, I.~P. \& Tyson, J.~A. 1996, ApJ, 473, L17

\bibitem[{Djorgovski} \& {Davis} 1987]{dd-87}
{Djorgovski}, S. \& {Davis}, M. 1987, \apj, 313, 59

\bibitem[{Dressler}, {Lynden-Bell}, {Burstein},  {Davies}, {Faber}, {Terlevich}, \& {Wegner} 1987]{dress-87}
{Dressler}, A., {Lynden-Bell}, D., {Burstein}, D., {Davies}, R.~L., {Faber},  S.~M., {Terlevich}, R., \& {Wegner}, G. 1987, \apj, 313, 42

\bibitem[{Faber} \& {Gallagher} 1979]{fg-77}
{Faber}, S.~M. \& {Gallagher}, J.~S. 1979, \araa, 17, 135

\bibitem[Faber \& Jackson 1976]{fj-76}
Faber, S.~M. \& Jackson, R.~E. 1976, ApJ, 204, 668

\bibitem[{Fischer}, {McKay}, {Sheldon}, {Connolly},  {Stebbins}, {Frieman}, {Jain}, {Joffre}, {Johnston}, {Bernstein}, {Annis},  {Bahcall}, {Brinkmann}, {Carr}, {Csabai}, {Gunn}, {Hennessy}, {Hindsley},  {Hull}, {Ivezi{\'c}}, {Knapp}, {Limmongkol}, {Lupton}, {Munn}, {Nash},  {Newberg}, {Owen}, {Pier}, {Rockosi}, {Schneider}, {Smith}, {Stoughton},  {Szalay}, {Szokoly}, {Thakar}, {Vogeley}, {Waddell}, {Weinberg}, {York}, \&  {The SDSS Collaboration} 2000]{fisch-00}
{Fischer}, P., {McKay}, T.~A., {Sheldon}, E., {Connolly}, A., {Stebbins}, A.,  {Frieman}, J.~A., {Jain}, B., {Joffre}, M., {et al.}, 2000, \aj, 120, 1198

\bibitem[{Folkes}, {Ronen}, {Price}, {Lahav},  {Colless}, {Maddox}, {Deeley}, {Glazebrook}, {Bland-Hawthorn}, {Cannon},  {Cole}, {Collins}, {Couch}, {Driver}, {Dalton}, {Efstathiou}, {Ellis},  {Frenk}, {Kaiser}, {Lewis}, {Lumsden}, {Peacock}, {Peterson}, {Sutherland},  \& {Taylor} 1999]{folkes-99}
{Folkes}, S., {Ronen}, S., {Price}, I., {Lahav}, O., {Colless}, M., {Maddox},  S., {Deeley}, K., {Glazebrook}, K., {et al.}, 1999, \mnras, 308, 459

\bibitem[Fukugita \& Turner 1991]{ft-91}
Fukugita, M. \& Turner, E.~L. 1991, MNRAS, 253, 99

\bibitem[Griffiths, Casertano, Im, \&  Ratnatunga 1996]{griff-96}
Griffiths, R.~E., Casertano, S., Im, M., \& Ratnatunga, K.~U. 1996, MNRAS, 282,  1159

\bibitem[{Guzman}, {Lucey}, \& {Bower} 1993]{glb-93}
{Guzman}, R., {Lucey}, J.~R., \& {Bower}, R.~G. 1993, \mnras, 265, 731

\bibitem[Hudson, Gwyn, Dahle, \& Kaiser 1998]{hud-98}
Hudson, M.~J., Gwyn, S. D.~J., Dahle, H., \& Kaiser, N. 1998, ApJ, 503, 531

\bibitem[{Jenkins}, {Frenk}, {White}, {Colberg},  {Cole}, {Evrard}, \& {Yoshida} 2001]{jenk-01}
{Jenkins}, A., {Frenk}, C.~S., {White}, S. D.~M., {Colberg}, J.~M., {Cole}, S.,  {Evrard}, A.~E., \& {Yoshida}, N. 2001, preprint (astro-ph/0005260)

\bibitem[{Jing}, {Mo}, \& {Boerner} 1998]{jmb-98}
{Jing}, Y.~P., {Mo}, H.~J., \& {Boerner}, G. 1998, \apj, 494, 1

\bibitem[Kaiser 1991]{kaiswk-91}
Kaiser, N. 1991, in Proc. New Insights into the Universe, Lecture Notes in  Physics 408, Val$\grave{e}$ncia, ed. V.~J. Martinez, M.~Portilla, \& D.~Saez  (Springer-Verlag), 279

\bibitem[{Kaiser} 2000]{kais-00}
{Kaiser}, N. 2000, \apj, 537, 555

\bibitem[Kaiser, Squires, Fahlman, \& Woods 1995]{ksfw-95}
Kaiser, N., Squires, G., Fahlman, G., \& Woods, D. 1995, in Proc. Meribel  conference, Clusters of Galaxies, ed. F.~Durret, A.~Mazure, \& J.~T.~T. Van  (Editions Fronti$\grave{e}$res, Gif-sur-Yvette)

\bibitem[Kaiser, Wilson, \&  Luppino 2001a]{kwl-01}
Kaiser, N., Wilson, G., \& Luppino, G. 2001a, paper I, cosmic  shear paper, preprint (astro-ph/0003338), KWL

\bibitem[Kaiser, Wilson, Luppino, \&  Dahle 2001b]{kwld-01}
Kaiser, N., Wilson, G., Luppino, G., \& Dahle, H. 2001b, preprint  (astro-ph/9907229), KWLD

\bibitem[Kaiser, Wilson, Luppino, Kofman,  Gioia, Metzger, \& Dahle 2001c]{kwlkgmd-01}
Kaiser, N., Wilson, G., Luppino, G., Kofman, L., Gioia, I., Metzger, M., \&  Dahle, H. 2001c, preprint (astro-ph/9809268), KWLKGMD

\bibitem[{Kim} \& {Fabbiano} 1995]{kf-95}
{Kim}, D. \& {Fabbiano}, G. 1995, \apj, 441, 182

\bibitem[{Kovner} \& {Milgrom} 1987]{km-87}
{Kovner}, I. \& {Milgrom}, M. 1987, \apjl, 321, L113

\bibitem[{Lilly}, {Tresse}, {Hammer}, {Crampton}, \& {Le  Fevre} 1995]{lilly-95}
{Lilly}, S.~J., {Tresse}, L., {Hammer}, F., {Crampton}, D., \& {Le Fevre}, O.  1995, \apj, 455, 108

\bibitem[{Mellier} 1999]{mel-99}
{Mellier}, Y. 1999, \araa, 37, 127

\bibitem[{Mushotzky}, {Loewenstein}, {Awaki},  {Makishima}, {Matsushita}, \& {Matsumoto} 1994]{mush-94}
{Mushotzky}, R.~F., {Loewenstein}, M., {Awaki}, H., {Makishima}, K.,  {Matsushita}, K., \& {Matsumoto}, H. 1994, \apjl, 436, L79

\bibitem[{Natarajan}, {Kneib}, {Smail}, \&  {Ellis} 1998]{nkse-98}
{Natarajan}, P., {Kneib}, J., {Smail}, I., \& {Ellis}, R.~S. 1998, \apj, 499,  600

\bibitem[{Navarro}, {Frenk}, \& {White} 1997]{nfw-97}
{Navarro}, J.~F., {Frenk}, C.~S., \& {White}, S. D.~M. 1997, \apj, 490, 493

\bibitem[{Trimble} 1987]{trim-87}
{Trimble}, V. 1987, \araa, 25, 425

\bibitem[{Trinchieri}, {Fabbiano}, \&  {Kim} 1997]{trinch-97}
{Trinchieri}, G., {Fabbiano}, G., \& {Kim}, D. 1997, \aap, 318, 361

\bibitem[{Trinchieri}, {Kim}, {Fabbiano}, \&  {Canizares} 1994]{trinch-94}
{Trinchieri}, G., {Kim}, D., {Fabbiano}, G., \& {Canizares}, C. R.~C. 1994,  \apj, 428, 555

\bibitem[{Tully} \& {Fisher} 1977]{tf-77}
{Tully}, R.~B. \& {Fisher}, J.~R. 1977, \aap, 54, 661

\bibitem[{Turner} 1976]{tur-76}
{Turner}, E.~L. 1976, \apj, 208, 304

\bibitem[Tyson, Valdes, Jarvis, \& Mills 1984]{tyson-84}
Tyson, J.~A., Valdes, F., Jarvis, J.~F., \& Mills, A.~P. 1984, ApJ, 281, L59

\bibitem[Wilson, Cowie, Barger, \&  Burke 2001a]{wcbb-01}
Wilson, G., Cowie, L.~L., Barger, A.~J., \& Burke, D.~J. 2001a,  star formation rates in the Hawaii Survey Fields, in prep.

\bibitem[Wilson \& Kaiser 2001]{wkcats-01}
Wilson, G. \& Kaiser, N. 2001, paper V, UH8K catalog paper, in prep

\bibitem[Wilson, Kaiser, \&  Luppino 2001b]{wkl-01}
Wilson, G., Kaiser, N., \& Luppino, G. 2001b, paper III, mass and  light paper, ApJ, in press, (astro-ph/0102396), WKL

\bibitem[{Wittman}, {Tyson}, {Kirkman},  {Dell'Antonio}, \& {Bernstein} 2000]{witt-00}
{Wittman}, D.~M., {Tyson}, J.~A., {Kirkman}, D., {Dell'Antonio}, I., \&  {Bernstein}, G. 2000, \nat, 405, 143

\bibitem[{Zaritsky}, {Smith}, {Frenk}, \&  {White} 1997]{zar-97}
{Zaritsky}, D., {Smith}, R., {Frenk}, C., \& {White}, S. D.~M. 1997, \apj, 478,  39

\end{thebibliography}
\end{document}